\documentclass{article}
\usepackage[preprint]{neurips_2026}

\usepackage[utf8]{inputenc} 
\usepackage[T1]{fontenc}    
\usepackage{hyperref}       
\usepackage{url}            
\usepackage{booktabs}       
\usepackage{amsfonts}       
\usepackage{nicefrac}       
\usepackage{microtype}      
\usepackage{xcolor}         
\usepackage{graphicx}
\usepackage{tabularx,array}
\usepackage{adjustbox}
\usepackage{booktabs}    
\usepackage{makecell}    
\usepackage{pifont}      
\usepackage[table]{xcolor}  
\usepackage{wrapfig}
\usepackage{graphicx}
\usepackage[nointegrals]{wasysym}
\usepackage{adjustbox}
\usepackage{rotating}

\newcommand{\bcircle}{$\newmoon$}
\newcommand{\wcircle}{$\LEFTcircle$}
\newcommand{\ecircle}{$\fullmoon$}
\usepackage{subcaption}
\usepackage{tikz}
\usetikzlibrary{arrows.meta, positioning, fit, calc}

\usepackage{amsmath,amsfonts,bm}

\def\eqref#1{equation~\ref{#1}}

\def\1{\bm{1}}

\DeclareMathAlphabet{\mathsfit}{\encodingdefault}{\sfdefault}{m}{sl}
\SetMathAlphabet{\mathsfit}{bold}{\encodingdefault}{\sfdefault}{bx}{n}

\usepackage{lscape}
\usepackage{pifont}
\usepackage{xspace}

\newcommand\tldrDone[1]{}

\newcommand\eg{e.g.\xspace}
\newcommand\ie{i.e.\xspace}

\newcommand\projectname{\textsc{EconEvals}\xspace}

\newcommand{\occupationcount}{1{,}016\xspace}
\newcommand{\dwacount}{2{,}087\xspace}
\newcommand{\taskcount}{18{,}796\xspace}
\newcommand{\softwareengineersoc}{15-1252.00\xspace}
\newcommand{\softwareengineertaskcount}{17\xspace}

\newcommand{\anthrophictaskshare}{17.2\%\xspace}
\newcommand{\anthropicsuppressionthreshold}{15 conversations or 5 unique accounts\xspace}
\newcommand{\gdpvaloccupationshare}{5\%\xspace}
\newcommand{\numbenchmarks}{226\xspace}
\newcommand{\nummodels}{10\xspace}
\newcommand{\lowUsageThreshold}{0.0025\%\xspace}

\newcommand{\realquerycount}{4{,}499{,}105\xspace}
\newcommand{\realconversations}{4{,}499{,}105\xspace}
\newcommand{\realDWAcount}{143\xspace}
\newcommand{\realDWAshare}{6.9\%\xspace}

\newcommand{\percentageOfRetrievedInUsage}{95.8\%\xspace} 

\newcommand{\percentageOfInUsageThatsRetrieved}{12.5\%\xspace} 

\newcommand{\totalUsageForInUsageAndRetrievedAbs}{27.4\%\xspace} 

\newcommand{\totalUsageForInUsageAndRetrieved}{31.3\%\xspace} 

\newcommand{\dwasOutsideOfSelected}{65.2\xspace} 

\newcommand{\pRetrieveAtThousandButNotThreeHundred}{0.09\xspace} 

\newcommand{\pRetrieveAtTenThousandButNotThreeHundred}{0.319\xspace}

\newcommand{\exprAdditionalDWAsFromGreaterDepth}{$(200 - 143) \times 0.319 = 18.18$\xspace}

\newcommand{\exprAdditionalDWAsFromGreaterBreadth}{$(1882 - 65.20) \times 0.319 = 579.56$\xspace}

\newcommand{\exprAllMissedDWAs}{$18.18 + 579.56 + 65.20 = 662.94$\xspace}

\newcommand{\realDWAupperbound}{806\xspace}

\newcommand{\realDWAshareupperbound}{38.6\%\xspace}

\newcommand{\realDWAOverallRecall}{0.18\xspace}

\newcommand{\percentOccupationsSaveHalfTasks}{46.6\%\xspace} 
\newcommand{\percentOccupationsSaveHalfTasksRound}{47\%\xspace} 

\newcommand{\percentOccupationsSaveHalfTasksHL}{68.7\%\xspace}

\newcommand{\percentOfExposedThatsLowUsage}{79.4\%\xspace} 
\newcommand{\percentOfExposedThatsLowUsageRound}{79\%\xspace} 
\newcommand{\exposedThatsLowUsageNum}{5681\xspace} 
\newcommand{\exposedThatsLowUsageDenom}{7159\xspace} 

\newcommand{\percentOfHighUsageThatsExposed}{73.3\%\xspace} 
\newcommand{\highUsageExposedNum}{1478\xspace} 
\newcommand{\highUsageExposedDenom}{2017\xspace} 

\newcommand{\exposureIntensityNumOcc}{857\xspace} 
\newcommand{\exposureIntensityPropOcc}{93\%\xspace} 

\newcommand{\realTotalSigFirst}{37\xspace} 

\newcommand{\realGPTOSSSigFirst}{35\xspace} 

\newcommand{\dwaRealSynthCorr}{0.67\xspace} 

\newcommand{\dwaSynthCorr}{0.86\xspace} 

\newcommand{\occGDPvalSynthCorr}{0.67\xspace} 

  {\begin{quote}\small
   \textsc{Simulated worker --- #1.}\par
   \textit{Task: #2}\par\smallskip}
  {\end{quote}}

\usepackage[most]{tcolorbox}

\definecolor{amberbg}{HTML}{FAEEDA}
\definecolor{ambertx}{HTML}{633806}
\definecolor{tealbg}{HTML}{E1F5EE}
\definecolor{tealtx}{HTML}{04342C}
\definecolor{cardbg}{HTML}{F5F4EE}
\definecolor{textmuted}{gray}{0.45}
\definecolor{textfaint}{gray}{0.65}

\newtcbox{\rubricbadge}{on line, arc=3pt, colback=amberbg, colframe=amberbg,
  boxrule=0pt, boxsep=0pt, top=2.5pt, bottom=2.5pt, left=6pt, right=6pt,
  fontupper=\color{ambertx}\bfseries\small}

\newtcbox{\simbadge}{on line, arc=3pt, colback=tealbg, colframe=tealbg,
  boxrule=0pt, boxsep=0pt, top=2.5pt, bottom=2.5pt, left=6pt, right=6pt,
  fontupper=\color{tealtx}\bfseries\small}

\def\Snospace~{\S{}}

\title{Economic Evaluations of Language Models}

\author{
Alexander Wan\\
Stanford University\\
\And
Stephane Hatgis-Kessell*\\
Stanford University\\
\And
Tomás Aguirre* \\
University of São Paulo \\
\And
Percy Liang\\
Stanford University\\
\And
Rishi Bommasani\\
Stanford University\\
}

\begin{document}

\maketitle
\begin{abstract}
Language models perform economically valuable work, yet they are not currently assessed for how well they perform every economically valuable task. We introduce \projectname as an open-source evaluation suite to measure capabilities relevant to tasks, work activities, and occupations in the US labor economy. 
We ground the evaluation suite in real user queries to language models where possible, and supplement these with synthetic data. 
Our evaluations improve coverage over OpenAI's GDPval benchmark, which is the existing state-of-the-art that  covers \gdpvaloccupationshare of US occupations, at $500\times$ lower cost.
Alongside benchmarks, we also introduce a \emph{simulation-based exposure measure} to estimate how much time current language model capabilities could save across all tasks belonging to all US occupations, with detailed accounting for each estimate. Our estimates indicate that current models could save workers substantial time on at least half of their tasks in $\percentOccupationsSaveHalfTasksRound$ of occupations. However, for $\percentOfExposedThatsLowUsageRound$ of tasks where we predict substantial time savings, observed Claude usage is low, suggesting that existing usage lags potential. 
Beyond inherent constraints of language model chatbots, our data identifies privacy and proprietary systems as the principal bottlenecks limiting further time savings from AI. 
Overall, we introduce adaptable infrastructure that grounds inferences about language models' labor-market impact in their current capabilities, which can be continually updated as capabilities improve.
\end{abstract}
\section{Introduction}
The economic impact of frontier AI is highly uncertain.
The public identifies job disruption as a top priority in relation to AI \citep{ipsos2025predictions, undp2025aihuman, aiindex2025publicopinion}.
Leaders across industry, computer science, economics, and government disagree in their economic forecasts \citep{whitehouse2026aiGreatDivergence, leap2025, karger2026forecasting}.  
For example, Nobel Laureate economist Daron Acemoglu predicts a modest cumulative GDP increase of roughly 0.9--1.6\% over the next decade from AI \citep{acemoglu2024simple}, while Anthropic CEO Dario Amodei predicts that AI could enable 10--20\% sustained annual GDP growth \citep{amodei2026adolescence}.
Existing evidence is limited \citep[see][]{chandar2025ailabormarkets} and facially provides contradictory accounts: several micro studies show that frontier AI meaningfully increases productivity, a few micro studies show it does not, and overall macro productivity evidence is fairly muted \citep[see][]{imas2026impact, delriochanona2025aijobsreviewtheory}. 

The uncertainty about the economic impacts of frontier AI conflicts with the near certainty of their improving capabilities.
Language models (LMs) saturate challenging benchmarks, often very quickly \citep{bengio2026aisafety, akhtar2026benchmarks, aiindex2026}.
This rapid progress has prompted the development of new evaluations that are indexed to ever-increasing human task complexity \citep{kwa2025measuring, metr2026timehorizons} or real-world utility \citep{patwardhan2025gdpval, mazeika2025remotelaborindex, vidgen2025apex, sun2026agentsexam}.
We posit that a central reason for unclear economic impact despite clear model capabilities is the misalignment between AI benchmarks and economic tasks.
\citet{wang2026doesagentdevelopmentreflect} quantifies this distribution shift: existing benchmarks are concentrated on topics like math and coding (39.7\% of benchmarking effort), even though the related occupations are a small share of the labor economy (3.5\% of U.S. jobs).\footnote{Many other organizational and regulatory factors also separate capabilities from impacts \citep{narayanan2025ainormaltechnology}.}

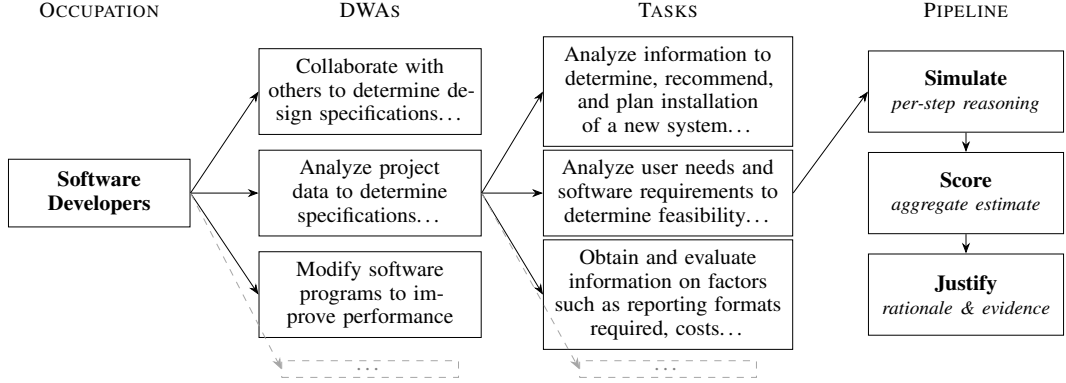
\begin{figure}[t]
\centering
\small
\resizebox{\linewidth}{!}{%
\begin{tikzpicture}[
    box/.style={rectangle, draw=black, line width=0.4pt, fill=white,
                align=center, inner sep=4pt, minimum height=0.75cm},
    occ/.style ={box, text width=2.4cm, minimum height=1.0cm},
    dwa/.style ={box, text width=3.0cm},
    task/.style={box, text width=3.4cm},
    pipe/.style={box, text width=2.6cm, minimum height=1.2cm},
    fade/.style={rectangle, draw=black!35, line width=0.3pt, dashed, fill=white,
                 align=center, inner sep=3pt, font=\itshape, text=black!50},
    arr/.style    ={-{Stealth[length=1.6mm]}, line width=0.4pt},
    arrfade/.style={-{Stealth[length=1.6mm]}, line width=0.3pt, black!35, dashed},
    hdr/.style={font=\scshape\small}
]

\node[hdr] at (0,    3.2) {Occupation};
\node[hdr] at (4,    3.2) {DWAs};
\node[hdr] at (8.4,  3.2) {Tasks};
\node[hdr] at (12.8, 3.2) {Pipeline};

\node[occ] (occ) at (0, 0.5) {\textbf{Software}\\\textbf{Developers}};

\node[dwa]  (d1) at (4, 2.0)  {Collaborate with others to determine design specifications\ldots};
\node[dwa]  (d2) at (4, 0.5)  {Analyze project data to determine specifications\ldots};
\node[dwa]  (d3) at (4, -1.0) {Modify software programs to improve performance};
\node[fade, text width=2.4cm] (dF) at (4, -2.1) {\ldots};

\node[task] (t1) at (8.4, 2.0)
  {Analyze information to determine, recommend, and plan installation of a new system\ldots};
\node[task] (t2) at (8.4, 0.5)
  {Analyze user needs and software requirements to determine feasibility\ldots};
\node[task] (t3) at (8.4, -1.0)
  {Obtain and evaluate information on factors such as reporting formats required, costs\ldots};
\node[fade, text width=2.4cm] (tF) at (8.4, -2.1) {\ldots};

\node[pipe] (p1) at (12.8, 2.0)  {\textbf{Simulate}\\\footnotesize\textit{per-step reasoning}};
\node[pipe] (p2) at (12.8, 0.5)  {\textbf{Score}\\\footnotesize\textit{aggregate estimate}};
\node[pipe] (p3) at (12.8, -1.0) {\textbf{Justify}\\\footnotesize\textit{rationale \& evidence}};

\draw[arr]     (occ.east) -- (d1.west);
\draw[arr]     (occ.east) -- (d2.west);
\draw[arr]     (occ.east) -- (d3.west);
\draw[arrfade] (occ.east) -- (dF.west);

\draw[arr]     (d2.east) -- (t1.west);
\draw[arr]     (d2.east) -- (t2.west);
\draw[arr]     (d2.east) -- (t3.west);
\draw[arrfade] (d2.east) -- (tF.west);

\draw[arr] (t2.east) -- (p1.west);

\draw[arr] (p1) -- (p2);
\draw[arr] (p2) -- (p3);

\end{tikzpicture}%
}
\caption{\textbf{O*NET work taxonomy and whitebox simulation-based task-level exposure estimates.} 
O*NET links occupations to DWAs and tasks as depicted for \textit{Software Developers} (15-1252.00).
Our method estimates task-level exposure across tasks using simulations and provides justifications.}
\label{fig:exposure-pipeline}
\end{figure}

We introduce the \projectname infrastructure to measure language model capabilities for work performed in the U.S. labor economy.\footnote{EconEvals is also used by \citet{fish2026econevalsbenchmarkslitmustests} to describe their research on AI's performance at making economic decisions.}
To codify work, we adopt a top-down approach using the Department of Labor's O*NET taxonomy.
O*NET maps the U.S. labor economy from \occupationcount\ occupations to \dwacount\ detailed work activities (DWAs) and \taskcount\ tasks.
For example, software engineers (SOC code \softwareengineersoc{}) perform \softwareengineertaskcount~tasks that include (i) analyzing user needs and software requirements to assess design feasibility, (ii) developing or directing software testing, validation, programming, or documentation, and (iii) modifying existing software to fix errors, adapt to new hardware, or improve performance (see \autoref{fig:exposure-pipeline}).

To measure the work capabilities of LMs, we ground our evaluations in real-world LM use.
While most LM usage data is not public, we use open-source datasets such as WildChat \citep{zhao2024wildchat} and LMSys \citep{zheng2023lmsyschat}.
Directly classifying the resulting \realquerycount\ real user queries into thousands of work categories would be very expensive, so we design a multi-stage retrieval pipeline to balance costs with pipeline precision and coverage.
This yields benchmark queries for \realDWAcount\ DWAs (\realDWAshare\ of all DWAs).
If, instead of balancing costs with coverage, we strictly maximized coverage, the upper bound on coverage from available public usage data would be \realDWAshareupperbound\ of all DWAs, which would increase costs by $300\times$. 
In comparison, the January 2026 Anthropic Economic Index reports Claude usage for only \anthrophictaskshare\ of tasks \citep{anthropic2026aeiv4}.

Given the limits of existing usage data, we design a synthetic data generation pipeline to improve query coverage of U.S. work, including for work categories where LMs are not yet adopted.
Specifically, we generate prompts at varying levels of coverage over a target task and use an LM verifier to check that the generated prompt both reflects what a worker in the occupation would do and contains all the information required to be answerable.
The result is benchmarks for all \dwacount\ DWAs, spanning all \occupationcount\ occupations in the U.S. labor economy.
This substantially improves coverage over past benchmarks including OpenAI's GDPval \citep{patwardhan2025gdpval} that covers 44 occupations ($<5\%$ of all U.S. occupations).
Our benchmarks are $500\times$ cheaper than GDPval, and we find that they predict occupational-level GDPval scores and are more predictive than generic capability benchmarks.

While benchmarks are the most common evaluation format in AI research, they foreground the disparities in model performance rather than how capabilities will yield economic value.
To complement the benchmarks we build, we also estimate \textit{exposure}: how much time would current capabilities save workers?
Exposure has become the dominant lens for translating between technological capabilities and economic returns \citep[\eg][]{autor2003skill}, including recent applications to frontier AI \citep[\eg][]{eloundou2024gpts}. 
Importantly, exposure guarantees neither increased labor productivity nor increased job displacement. 
We introduce a \emph{whitebox simulation-based exposure measure} that builds on prior work using LMs to simulate human behavior \citep{park2023generative}.
Our exposure measure simulates a worker using a LM chatbot to complete a task, producing a detailed accounting of the steps involved in performing the task, the baseline time per step, and the time savings per step attributable to LMs.
This satisfies several desiderata unaddressed by previous measures (\autoref{tab:exposure-measures-summary}).

Using our simulation-based exposure measure, we find that current LMs could save workers substantial time on at least half of the tasks in $\percentOccupationsSaveHalfTasks$ of U.S. occupations.
In spite of the broad potential for substantial time savings, we find that current usage does not cover all of these task-level opportunities: $\percentOfExposedThatsLowUsage$ of tasks that are predicted to be exposed by our metric have low Claude usage according to Anthropic's statistics.
Further, by not only producing quantitative exposure estimates but an underlying rationale for how these time savings are achieved, we identify bottlenecks that inhibit technological capabilities from translating to productivity gains. 
Beyond inherent constraints of LMs like physical interaction, privacy constraints and proprietary systems are the primary current bottlenecks inhibiting further task-level time savings. 
We hope that by releasing all of our data, including the underlying instances of simulated workers interacting with language models, we can directly support real human workers in discovering how to better derive work benefits from current language models.
Overall, our infrastructure helps bridge the gap between AI capabilities and labor market impact, and is designed to keep pace with both technological change (\eg new models) and economic change (\eg new tasks). 

\begin{table}[t]
  \centering
  \footnotesize
  \setlength{\tabcolsep}{6pt}
  \renewcommand{\arraystretch}{1.2}
  \begin{tabular}{lcccc}
    \toprule
    \textbf{Approach} & \makecell{\textbf{Open  }\\\textbf{data}} & \makecell{\textbf{Grounded in task-}\\\textbf{execution performance}} & \makecell{\textbf{Comparable}\\\textbf{across models}} & \makecell{\textbf{Readily extensible}\\\textbf{to the full task space}}\\
    \midrule
    Rubric-based exposure & \bcircle & \ecircle & \wcircle & \bcircle\\
    \rowcolor[gray]{0.9}
    Adoption-based exposure & \ecircle & \bcircle & \ecircle & \wcircle\\
    Economically relevant benchmarks & \wcircle & \bcircle & \bcircle & \ecircle\\
    \rowcolor[gray]{0.9}
    \textbf{Simulation-based exposure (ours)} & \bcircle & \bcircle & \bcircle & \bcircle\\
    \bottomrule
  \end{tabular}
  \vspace{2mm}
  \caption{\textbf{Comparing approaches to exposure estimation for four desiderata.} \bcircle{} indicates the property is fully satisfied; \wcircle{} indicates partially satisfied; \ecircle{} indicates not satisfied.}
  \label{tab:exposure-measures-summary}
\end{table}
\section{Data}
\label{sec:data}
\projectname evaluations are grounded in real usage where possible.
Given the limited coverage of US work in current (public) usage, we ensure complete coverage through synthetic data.

\subsection{Real Data}
We accumulate real user conversations from public datasets and map these conversations to work categories through a multi-step pipeline involving embedding-based retrieval and language model classifiers.
Subject to cost constraints, we optimize the accuracy of the pipeline.
To assess the overall quality, we report (i) total costs, (ii) precision in mapping from conversations to work categories, and (iii) coverage of the space of work categories.

\paragraph{Public datasets.}
We draw on five large public corpora of real user--chatbot conversations published with user consent that total to $\realconversations$ conversations.
WildChat-4.8M \citep{zhao2024wildchat} provides 3.2M conversations from free GPT-3.5-Turbo and GPT-4 chat services hosted on Hugging Face Spaces.
LMSYS-Chat-1M \citep{zheng2023lmsyschat} provides 1M conversations from the Vicuna demo and Chatbot Arena.
\citet{chiang2024chatbotarena} provides three releases (55k, 100k, 140k) of Chatbot Arena human preference data.

\paragraph{Pipeline design.}
We map user conversations to work categories at the detailed work activity (DWA) level of the O*NET taxonomy.
We select this level of the hierarchy as the most fine-grained level of the O*NET taxonomy that was feasible given our data.\footnote{Tasks are more fine-grained than DWAs but in initial experiments we found tasks to be infeasible: pipeline precision was considerably lower and the number of identified conversations per task to be too low for our purposes.}
Directly classifying $\realconversations$ into $\dwacount$ categories would be prohibitively costly.
Therefore, we decompose the classification problem into an initial lightweight retrieval phase followed by a more costly classification phase that only triggers for a much smaller subset of (conversation, DWA) pairs. However, to ensure adequate coverage over occupations, we retrieve at the more fine-grained task level where each task corresponds to a single occupation (i.e., for each DWA, we perform retrieval \& pairwise classification for all tasks associated with that DWA). Importantly, the benchmarks we produce are still at the DWA-level (we try to retrieve at least 50 samples per DWA); we just use task-level metadata to improve the coverage of our pipeline.

To implement the retrieval phase, we use standard embedding-based retrieval methods \citep{douze2024faiss}. We embed every conversation as well as the tasks/occupations associated with each DWA.
Since classifying all $9.4 \times 10^{9}$ DWA--conversation pairs (or the even larger number of task--conversation pairs) is infeasible, we reduce cost by selecting promising DWAs that are likely to have enough conversational coverage. 
We select 200 DWAs that have the highest number of relevant conversations based on retrieving the top 10 most similar conversations for each associated task and then running the subsequent pairwise classification pipeline.
Then, for each selected DWA, we retrieve the top 300 most similar conversation for each associated task.

Given the retrieved (DWA, conversation) pairs, we classify them using hierarchical triage: we use cheap language models to narrow the initial pool and expensive language models to further refine it to manage costs while improving precision.
For all (DWA, query) pairs that survive, we filter out low-quality queries and non-work queries.
Finally, given the resulting high-quality work-specific conversations for each DWA, we convert them into benchmarking queries by using an LM to select the user-turn most relevant to the DWA.
The full pipeline implementation with each step is described in \autoref{app:real-pipeline-implementation}.

\paragraph{Quality.}
We designed our query-task mapping to maximize precision and minimize overall costs.
To estimate costs, we aggregate costs per pipeline step: the overall cost is approximately \$768 for nearly all tasks and occupations in O*NET
, which breaks down into \$242 for embedding the data/queries, \$477 for multi-stage language model classification, and \$67 for subsequent data filtering. To estimate precision, we label a sample of 50 query-task pairs identified by our pipeline based on whether we assess the query-task mapping to be correct: the overall pipeline precision is 0.91.

In addition to high precision and low costs, our third desiderata is to maximize recall.
Since we perform retrieval and do not classify every possible query-task pair, measuring recall is more complicated than measuring precision.
In \autoref{app:real-pipeline-evaluation}, we plot the results of two experiments that test how DWA-level coverage would improve with (i) broader retrieval beyond the selected 200 DWAs and (ii) deeper retrieval beyond the the top-300 most similar retrieved queries.
The overall pipeline recall is \realDWAOverallRecall: the pipeline identifies \realDWAcount DWAs with 50+ queries and the experiments predict that sufficiently broad and deep retrieval would identify \realDWAupperbound DWAs with 50+ queries.
This recall is achieved at a cost savings of approximately $300\times$ since we only perform the costly language model classification step for $300$ queries $\times~200$ DWAs instead of $10000$ queries $\times~2000$ DWAs.
We prioritize recall to improve task coverage given the limited task coverage of existing benchmarks \citep{wang2026doesagentdevelopmentreflect}.
For the criterion of at least 50 queries per covered DWA, the pipeline covers \realDWAshare of O*NET DWAs and is upper bound at \realDWAshareupperbound given our recall estimate. 

To contextualize our task coverage, we consider usage statistics published by Anthropic based on their proprietary Claude usage data.
The January 2026 Anthropic Economic Index publishes usage for tasks with at least \anthropicsuppressionthreshold: \anthrophictaskshare of tasks and roughly half of the DWAs are covered.\footnote{These values are for O*NET 20.1 and are at the task level: we say a DWA is covered if any associated task is covered.}
Our retrieval is high precision but low recall: \percentageOfRetrievedInUsage of the DWAs for which we retrieved samples for appear in the Economic Index usage data, but our retrieval only covers \percentageOfInUsageThatsRetrieved of the their DWAs, which accounts for \totalUsageForInUsageAndRetrieved of their usage.\footnote{The share of usage covered by our retrieved DWAs reported above, \totalUsageForInUsageAndRetrieved, is normalized against the Claude usage that (1)~the Economic Index is able to classify into a task category and (2)~can be mapped to DWAs present in both O*NET 20.1 and 29.2. The unnormalized value, the percentage of total Claude usage covered by our retrieved DWAs, is \totalUsageForInUsageAndRetrievedAbs.}

For example, we find that their most frequent DWAs (``Modify software programs to improve performance'' and ``Tutor students who need extra assistance'') are not retrieved but similar  education and programming DWAs are covered like ``Write computer programming code'' and ``Develop instructional materials''.
Ultimately, current usage data of all types does not cover the full diversity of U.S. work.

\subsection{Synthetic User Queries}
Public usage data has limited work coverage and even proprietary usage is fundamentally constrained to how AI is currently used.
However, we believe AI should be evaluated for all work use cases: evaluations could inform the procurement and adoption of AI for new tasks as they demonstrate technological improvement.
We introduce a simulation-based synthetic data generation pipeline to cover (essentially) all U.S. work.\footnote{In initial experiments, we simply prompted LMs to generate queries but found the resulting synthetic data low quality.} 

To control for the complexity of the queries we generate, for each task and occupation, we first create a worker persona and then have \texttt{GPT-5-mini} roleplay as a worker with this persona responding to an interviewer asking about time savings.
To begin, we start with the maximal time savings and see if a query can be generated to justify such high time savings, iteratively backing off to lower and lower exposure estimates. 
Since high time savings may arise due to LM hallucinations or other departures from realistic economic modeling, we introduce additional verification steps.
The implementation for synthetic data generation is described in more detail in \autoref{app:synthetic-data}.

\begin{figure}[tbp]
    \centering
    \vspace{0.1cm}
    \begin{subfigure}[t]{0.49\linewidth}
        \centering
        \includegraphics[width=\linewidth]{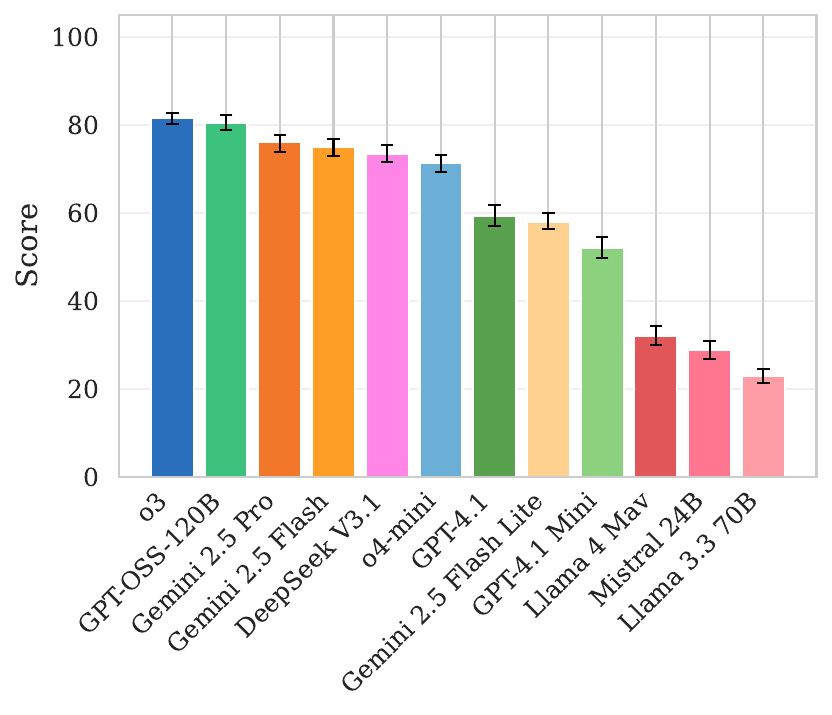}
        \label{fig:random_lb}
    \end{subfigure}
    \hfill
    \begin{subfigure}[t]{0.49\linewidth}
        \centering
        \includegraphics[width=\linewidth]{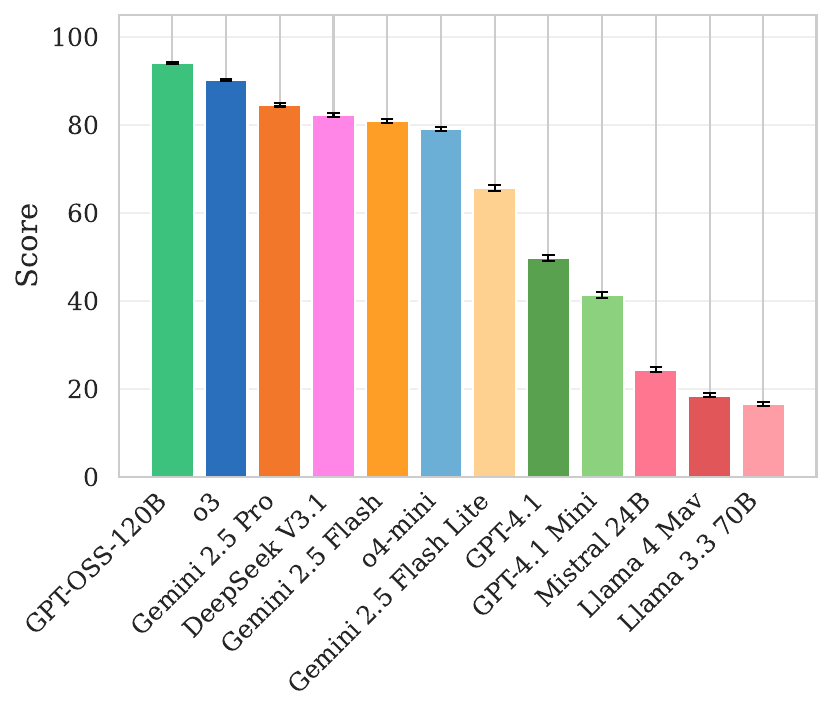}
        \label{fig:arena_lb}
    \end{subfigure}
    \vspace{-0.3cm}
    \caption{\textbf{Comparing economic benchmarks to raw public usage benchmarks.}
    The left subfigure depicts model performance on 500 randomly sampled public user queries whereas the right depicts model performance across all our \realDWAcount 50-instance DWA-level benchmarks.
    }
    \label{fig:random_vs_economic}
\end{figure}

\vspace{0.3cm}
\section{Benchmarks}
\label{sec:benchmarks}

We present results for \numbenchmarks benchmarks using our real and synthetic economically-categorized data as the benchmark queries and using language model judges to score model responses to these queries.

\paragraph{Scoring.}
We evaluate \nummodels models on our benchmarks, namely \texttt{gpt-4.1-mini}, \texttt{gpt-4.1}, \texttt{o4-mini}, \texttt{o3}, \texttt{gpt-oss-120b}, \texttt{gemini-2.5-pro}, \texttt{gemini-2.5-flash}, \texttt{gemini-2.5-flash-lite}, \texttt{Llama-3.3-70B}, \texttt{DeepSeek-V3.1}.\footnote{We bias towards cheaper models to reduce evaluation cost, so we do not report results for some expensive frontier models.} We cover different model sizes (e.g., Gemini 2.5 flash-lite, Gemini 2.5 flash, Gemini 2.5 Pro), different model developers (e.g., OpenAI, Google), flagship thinking/non-thinking models (e.g., o3 vs GPT-4.1), and prominent open/closed models (e.g., DeepSeek-V3.1 vs GPT-4.1).

Model responses are generated for each assessed model on each benchmark.
To assess the quality of model responses, we use a language model judge.
Following prior work \citep{li2024arenahardauto}, we elicit binary preferences from the judge: is the evaluated model's response better, worse, significantly better, significantly worse, or the same as the response of a reference model?\footnote{We set \texttt{o3-mini} as the reference model for analyzing the results in the paper.}
Overall, we present results for \numbenchmarks benchmarks in this work: \realDWAcount 50-instance DWA-level benchmarks based on real data, 40 50-instance DWA-level benchmarks based on synthetic data, and 43 50-instance occupation-level benchmarks for the occupations in GDPval \citep{patwardhan2025gdpval}.\footnote{GDPval has 44 occupations, but we remove Buyers and Purchasing Agents as it’s not present in O*NET 30.2.}

\paragraph{DWA-level benchmarks based on real data.}
In \autoref{fig:random_vs_economic}, we plot model performance in aggregate on public usage data as-is and across our per-DWA benchmarks.
While the general trends are similar, \texttt{gpt-oss-120B} performs the best by a clear margin on our economic benchmarks as compared to the raw usage where it is tied for first.
Disaggregating by DWA confirms this: of the \realTotalSigFirst benchmarks where a model is in first place by a statistically significant margin (\ie its confidence interval does not overlap with any other model's confidence interval), \realGPTOSSSigFirst have \texttt{gpt-oss-120B} in first place.
Overall, the per-DWA benchmarks generally induce similar rankings as the average ranking across all \realDWAcount DWAs.
Further, the variance in model performance significantly increases with our economic benchmarks, indicating that our economic weighting identifies greater separation in model quality compared to the raw public usage data.
This is particularly noteworthy since a sizable fraction of our usage data derives from the popular leaderboard LMArena.

\paragraph{DWA-level benchmarks based on synthetic data.}
Our pipeline for classifying public usage data amounts to DWA-level benchmarks for only about 7\% (\realDWAcount/\dwacount) of the DWAs.
Based on our synthetic data, we evaluate models on 40 DWAs: 20 randomly sampled from the \realDWAcount DWAs that the real data covers and 20 randomly sampled from the remaining 93\% of uncovered DWAs.
In \autoref{fig:synthetic-DWA}, we report the results for all 40 DWAs.\footnote{We do not evaluate Llama-4-Maverick and Mistral-24B as they were were deprecated on Together AI.}
On the 20 DWAs covered by the real data, we find the synthetic benchmark results, on average, have a spearman correlation of \dwaRealSynthCorr
with the real benchmark results (\autoref{fig:real-vs-synthetic-benchmarks}).
Comparing the synthetic benchmark results for the two sets of 20 DWAs, we find an average spearman correlation of \dwaSynthCorr (\autoref{fig:use-vs-nonuse-benchmarks}).

Disaggregating the results by DWA, we again find that the model ranking is similar across DWAs.
But DWAs differ in how much they separate models (\autoref{fig:synthetic-DWA-variance}).
DWAs that involve instructing or advising tend to have the largest variance in model performance and best separate model quality.
In contrast, DWAs that involve verification or record maintenance tend to have the least variance in model performance and thereby generally do not identify differences in model quality.
For example, all 3 DWAs that involve maintaining records (\ie ``Maintain medical records'', ``Maintain operational records'', ``Maintain the inventory of equipment'') are in the bottom quintile when ranking by variance whereas ``Provide technical guidance to other personnel'' is the DWA with the greatest variance.
The sole counterexample we find to these trends is the ``Document operational procedures'' detailed work activity, which resembles record maintenance, but is the third highest DWA by variance.

\paragraph{Occupation-level benchmarks based on synthetic data.}
\citet{patwardhan2025gdpval} built the GDPval benchmark using queries sourced from workers to yield benchmarks for 44 occupations, each containing 30 task instances. 
The selected occupations all predominantly perform knowledge work and belong to the 9 U.S. sectors that each contribute over 5\% of GDP, prioritizing the occupations with the largest total wage-and-compensation contribution within each sector.
Compared to GDPval, the \projectname synthetic benchmarks have five advantages: (i) greater coverage of U.S. work, (ii) clearer understanding of the task-level coverage within an occupation, (iii) more instances per occupation, (iv) lower cost to produce the benchmark, and (v) full transparency and open-source data. 
We estimate that our evaluation queries cost less than \$1000 to produce whereas the GDPval evaluation queries cost more than \$500,000 to produce (\autoref{app:benchmarks}).
However, GDPval has superior data quality based on the small subset of queries they publish: they better test frontier capabilities (\eg multi-turn, agents, tool use and web search) rather than our narrower focus on language models and they more directly align with work deliverables than task completion (\eg producing a legal brief rather than performing legal research).

We test whether our occupation-level benchmark scores predict GDPval scores for the same occupations.
We find that the spearman correlation between the synthetic and GDPval scores is, on average, \occGDPvalSynthCorr.
Since prior work finds that many capability benchmarks are correlated \citep{ho2025rosetta}, we further test whether generic capability measures also predict GDPval scores or whether our benchmarks provide increased explanatory power. Compared to the existing capability benchmarks, the generated synthetic benchmarks win or tie in terms of spearman correlation in the majority of cases. However, only seven out of nine are statistically significant (\autoref{fig:predicting-GDPval}).
\begin{figure}[ht]
    \centering
    \includegraphics[width=\textwidth]{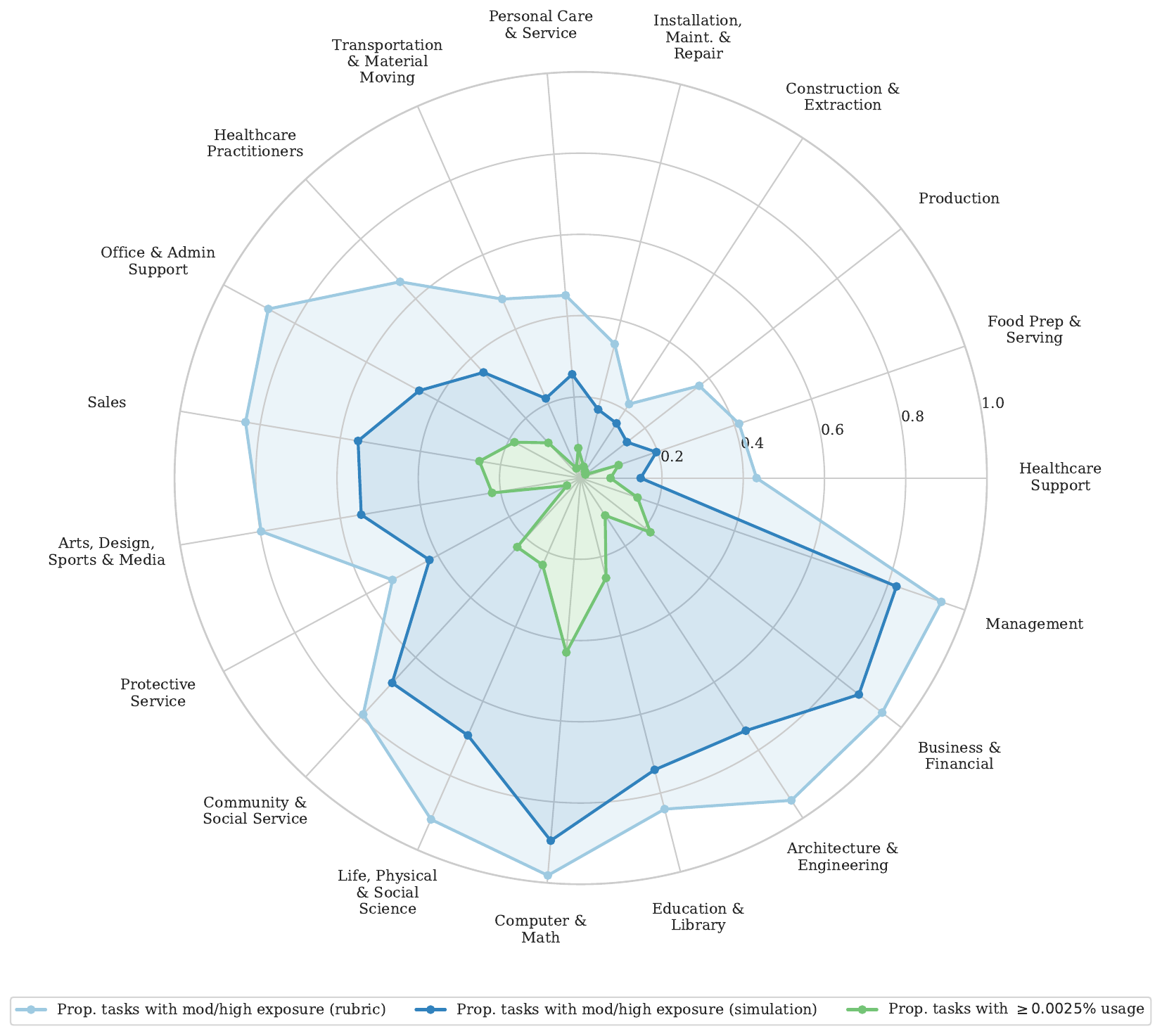}
    \caption{\textbf{Average proportion of moderate or highly exposed tasks versus average proportion of tasks with significant observed usage.} We average the proportion of tasks that are exposed/have usage for each occupation, then average the occupation-level values for each major group. Across all major groups, both exposure predictions exceed actual Claude usage, indicating that AI could save time on far more tasks than it is currently used for.}
    \label{fig:exposure-usage}
    \vspace{-5mm}
\end{figure}

\section{Exposure}
\label{sec:exposure}

We aim to understand how language model capabilities are transformed into impacts on the labor market.  
Economists have developed \textit{exposure measures} to study this transformation, which measure how much time workers would save given new technologies \citep{autor2003skill,acemoglu2011skills,acemoglu2019automation}.
Several works estimate AI-driven exposure \citep{felten2018method,webb2020impact,felten2021occupational,tolan2021measuring,felten2023occupational,pizzinelli2023labor,eloundou2024gpts,massenkoffmccrory2026labor} with \citet{eloundou2024gpts} popularizing exposure measurement for LMs. 

\paragraph{Scoring.}
Existing exposure estimates for LMs \citep{eloundou2024gpts, hosseini2026generative_entry_barriers} generally (i) specify a list of current capabilities and (ii) use LMs, workers, or occupational experts to annotate the extent to which these capabilities would save time for a given work task.
We call these \textit{blackbox rubric-based} exposure measures: the measure yields a task-level exposure score based on a predefined rubric of capabilities with no transparency into how the score was computed.
Exposure estimation is a complex task that requires jointly reasoning about general technological capabilities, specific work tasks, and how the two interact.
For example, to estimate language model exposure for the task of ``Taking customers' orders'' requires accounting for the fact that even if LMs can do certain steps like summing the prices in an order or making item recommendations, they would also need to handle synchronous in-person customer interactions to further save time.

We introduce the first \textit{whitebox simulation-based} exposure method (see \autoref{fig:exposure-comparison}).
To estimate the exposure of a task, we roleplay a ``worker'' using a language model and construct a decomposition of each task into steps.
Given this decomposition, the simulated ``worker'' provides a baseline estimate for the time per step and the amount this can be reduced given current language model capabilities.
The resulting per-step time savings are aggregated to determine the overall time savings as our exposure prediction (we run this simulation process three times and take the median time-savings to account for outliers).
Finally, we discretize these time savings into four categories, similar to prior work \citep{hosseini2026generative_entry_barriers}.
Through this simulation, we produce a numerical exposure estimate backed by an underlying reasoning trace that directly accounts for what we understand the task to entail and which steps within that task are accelerated by current language model capabilities.
We describe the full implementation of our simulation with examples of the resulting per-task exposure accounting in \autoref{app:detailed_exposure_description}. 

\begin{figure}[t]
\centering
\small
\begin{tcolorbox}[
  enhanced, colback=white, colframe=black, boxrule=0.4pt, arc=0pt,
  boxsep=2pt, left=10pt, right=10pt, top=8pt, bottom=8pt,
  width=\linewidth
]
\textbf{Occupation:} Hosts and Hostesses, Restaurant, Lounge, and Coffee Shop.\\
\textbf{Task:} Receive and record patrons' dining reservations.

\vspace{3pt}\hrule\vspace{5pt}

\noindent
\begin{minipage}[t]{0.475\linewidth}
\textbf{Blackbox rubric-based exposure.}

\smallskip
\textit{Prompt rubric (excerpt).}

\smallskip
\textsc{**T0 — No Automation**} System cannot perform any meaningful component of the task. Typically highly physical, deeply emotional, or restricted by legal requirements.

[\,\ldots\,]

\textsc{**T3 — High Automation (Human-in-the-Loop)**} System can perform 80–100\% at high quality, BUT human oversight is required because of: - liability/safety, - stakeholder trust expectations, - rare catastrophic failure modes

\textsc{**T4 — Full Automation (Autonomous)**} System performs 100\% of the task with high quality. Human oversight is *not routinely needed* and humans are *not liable* for errors. The entire workflow is digitally executable end-to-end.
\end{minipage}\hfill
\begin{minipage}[t]{0.475\linewidth}
\textbf{Whitebox simulation-based exposure.}

\smallskip
\textit{Per-step reasoning (excerpt).}

\smallskip
11) Updating waitlist / callbacks (1–10 min intermittently)\\
- What it takes now: add to waitlist and call people when spaces open; manual calls take minutes each.
- Chatbot effect: only useful if integrated to generate/send messages or summarize who's on the list. Standalone text generation is marginal; integrated automation would help a lot but isn't in the demo.\\
- Time saved: none to small without integration.
\end{minipage}

\vspace{6pt}\hrule\vspace{5pt}

\noindent
\begin{minipage}{0.475\linewidth}
\centering\textbf{Estimated time saved: 80--100\%}
\end{minipage}\hfill
\begin{minipage}{0.475\linewidth}
\centering\textbf{Estimated time saved: 0--25\%}
\end{minipage}

\end{tcolorbox}
\caption{\textbf{Prior blackbox rubric-based vs. our whitebox simulation-based exposure measures.} 
The left side shows an excerpt from the blackbox rubric-based measure from prior work \citep{hosseini2026generative_entry_barriers}.
The right side shows our decomposition of the task into steps with step-wise accounting to determine the overall time-savings based on our worker simulation.}
\label{fig:exposure-comparison}
\end{figure}

\paragraph{Comparing to LM Usage Data}
\begin{figure}[ht]
    \centering
    \includegraphics[width=\textwidth]{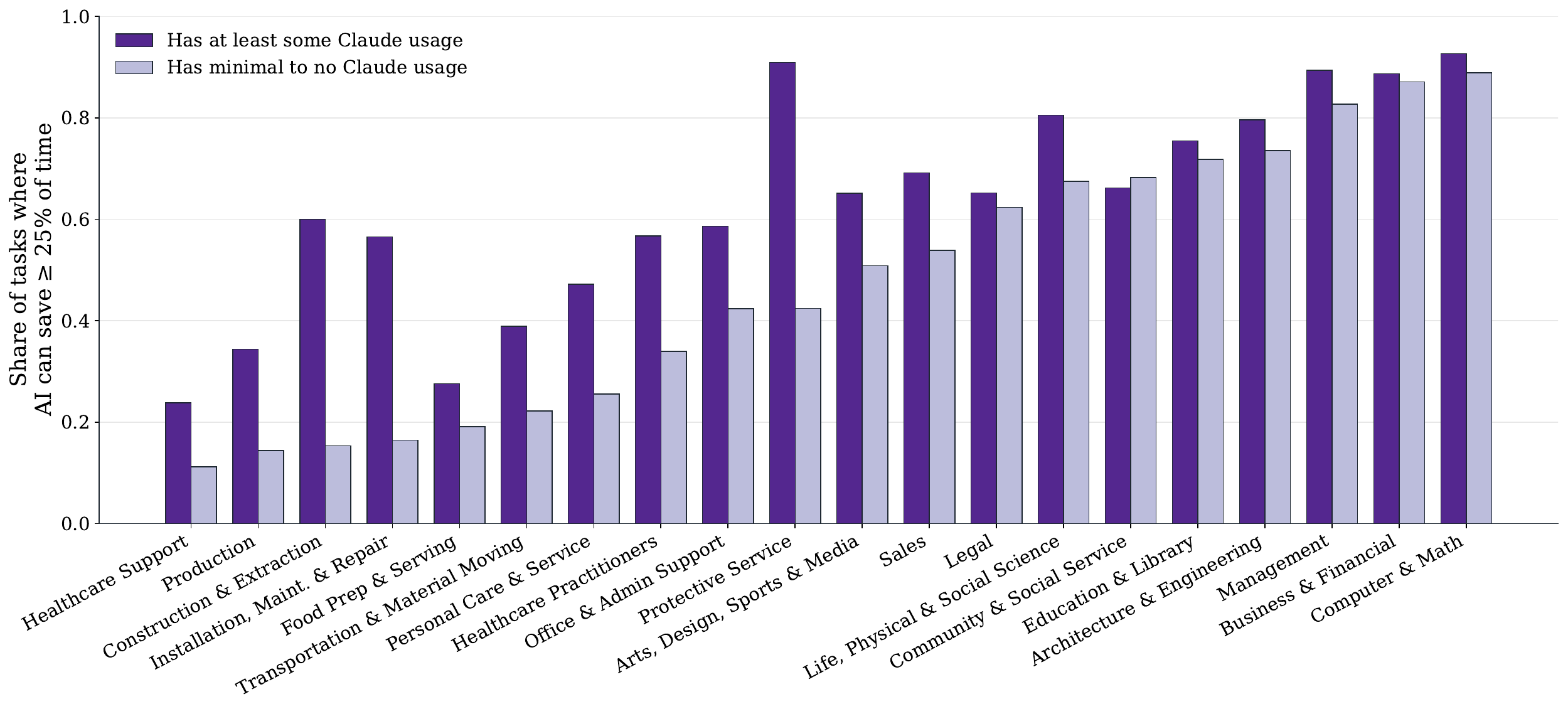}
    \caption{\textbf{Simulation-based exposure estimates by occupational group and Claude usage.}
    We plot the exposure estimate for each SOC major group aggregated across all tasks within the group.
    Results are stratified based on whether the underlying tasks have significant reported Claude usage (dark; at least \lowUsageThreshold of Anthropic's work-related Claude.ai traffic) or not (light).}
    \label{fig:exposure-usage}
    \vspace{-5mm}
\end{figure}

We compare our exposure estimates to the usage data from the Anthropic Economic Index \citep{handa2025economic}.
\autoref{fig:exposure-usage} compares the simulation-based exposure for tasks that appear in at least $\lowUsageThreshold$ of Anthropic's work related traffic for Claude.ai versus those that do not. Among tasks that see some Claude usage, \percentOfHighUsageThatsExposed (\highUsageExposedNum/\highUsageExposedDenom) are ones where AI can save at least 25\% of time, indicating that adoption is concentrated on tasks with meaningful exposure. However, among tasks where AI can save at least 25\% of time, \percentOfExposedThatsLowUsage (\exposedThatsLowUsageNum/\exposedThatsLowUsageDenom) have minimal to no Claude usage.

High predicted exposure scores for tasks with little to no current Claude usage suggest that LMs could be applied more extensively in people's jobs than current usage patterns indicate. Such gaps between capability and adoption are characteristic of general-purpose technologies, which historically diffuse through the economy only as firms develop the complementary workflows, skills, and organizational practices needed to deploy them effectively. For example, for Online Merchants calculating revenue and expenses, our exposure measure predicts that LMs could generate spreadsheet formulas, navigate platform tools like Amazon Seller Central, and reconcile bank statements against payment-processor records. For Architects administering construction contracts, LMs could summarize contract clauses, flag inconsistencies between schedules of values and inspection evidence, and produce post-meeting notes from contractor calls. Despite the predicted time-savings for these tasks and occupations under the simulation-based exposure measure, these tasks see little to no Claude usage. 
In \autoref{app:low_exposure_nonzero_usage} we further analyze tasks that the simulation-based-exposure measure labels as exposed to LMs but have minimal Claude usage, and tasks that have at least some Claude usage but are labeled as minimally exposed.

\begin{figure}[t]
    \centering
    \vspace{0.5cm}
    \includegraphics[width=\textwidth]{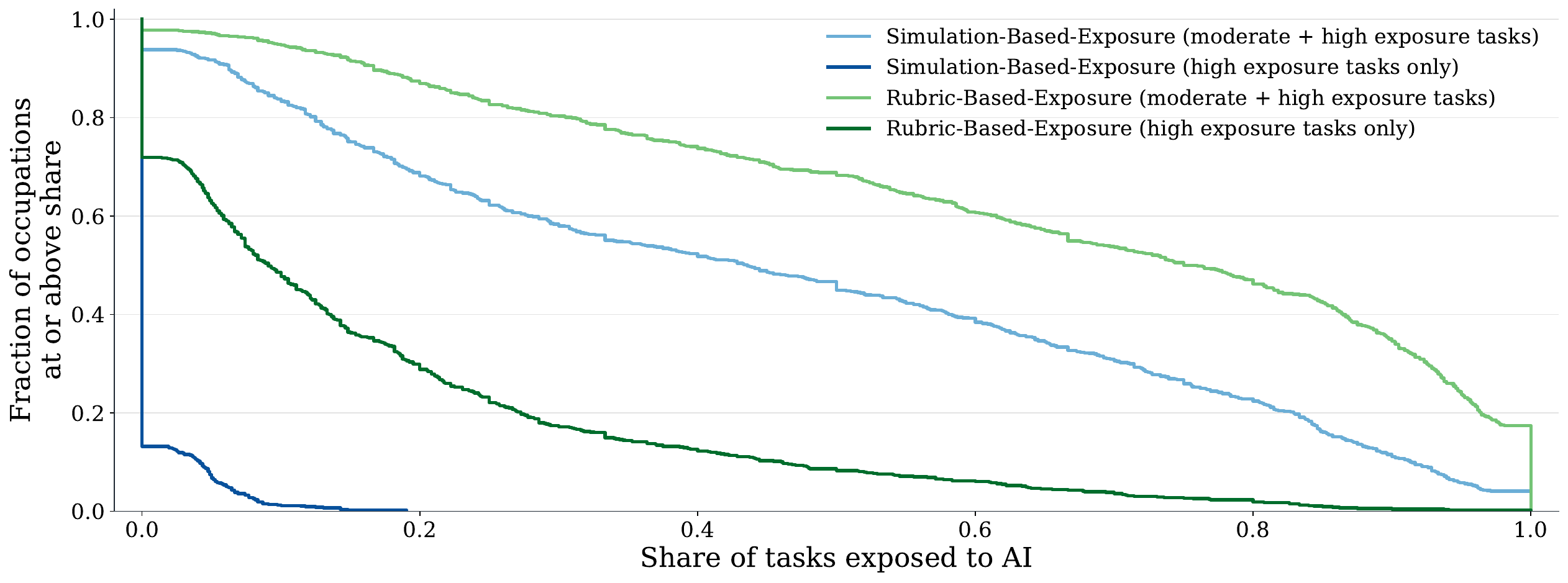}
    \caption{\textbf{Simulation-based vs. rubric-based task-level exposure estimates.}
    We plot the occupation-level exposure for our method and prior work as a function of the minimum proportion of tasks exposed. At both the high and moderate-to-high threshold, the simulation-based method predicts lower exposure as it surfaces bottlenecks to adoption that only arise from a more detailed accounting of AI-use.}
    \label{fig:exposure-intensity}
\end{figure}

\begin{figure}[t]
    \centering
    \begin{subfigure}[b]{0.49\textwidth}
        \centering
        \includegraphics[width=\textwidth,trim={0.3cm 0 0.4cm 0},clip]{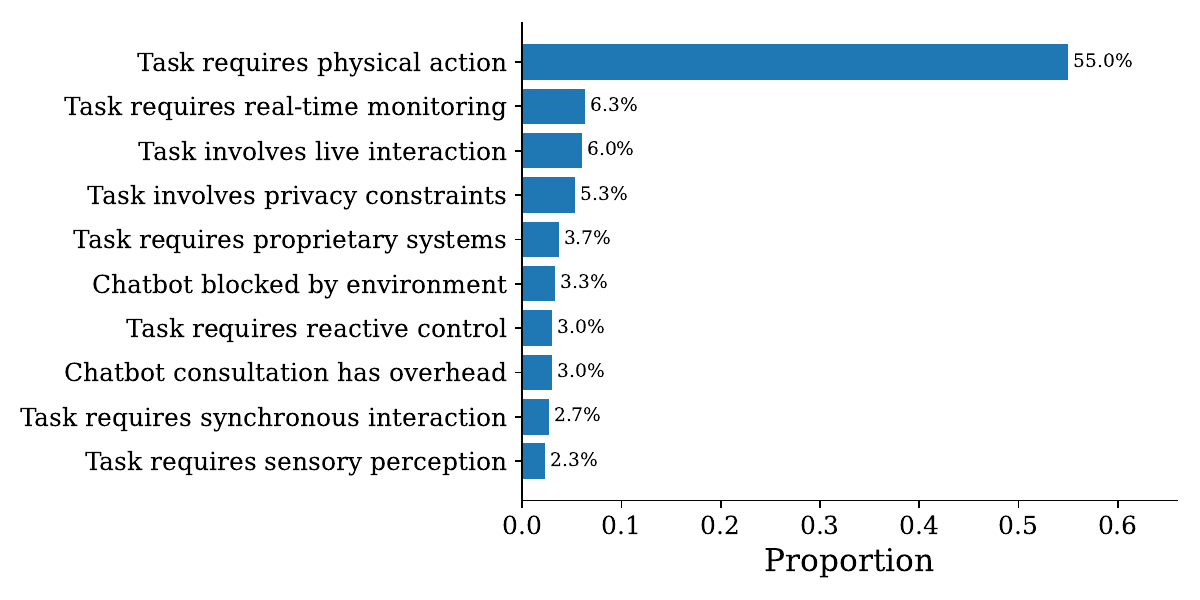}
        \label{fig:subfig-a}
    \end{subfigure}
    \hfill
    \begin{subfigure}[b]{0.49\textwidth}
        \centering
        \includegraphics[width=\textwidth,trim={0.3cm 0 0.4cm 0},clip]{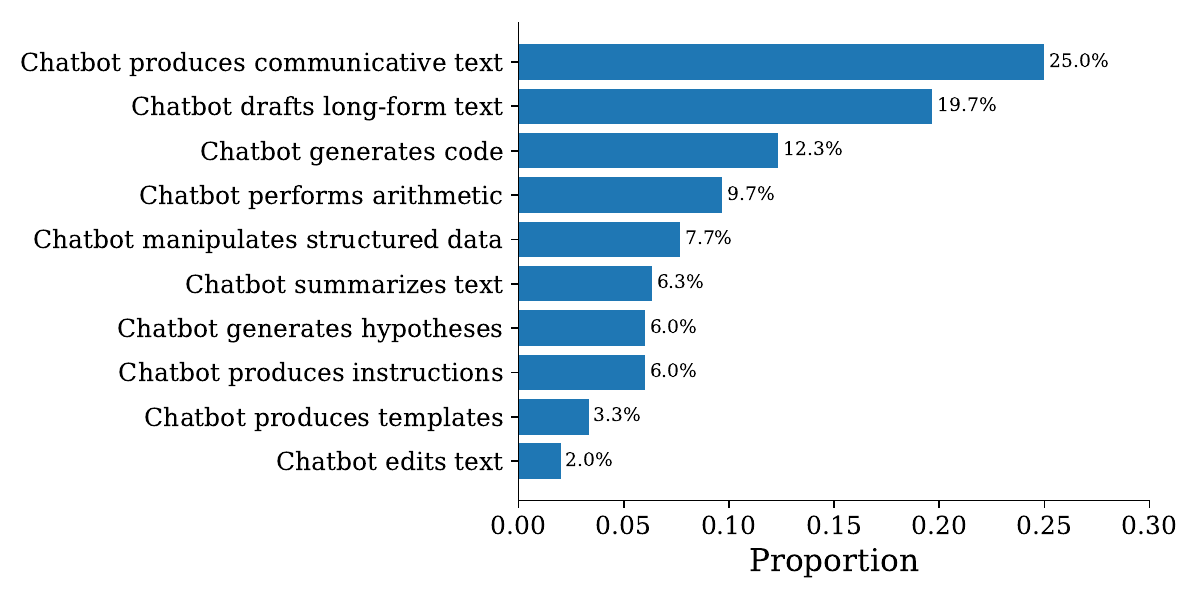}
        \label{fig:subfig-b}
    \end{subfigure}
\vspace{-0.5cm}
\caption{\textbf{Explaining time savings via reasoning trace analysis.}
The left side shows bottlenecks predicted by our simulation-based exposure predictions that limit time savings below $25\%$. 
The right side shows the predicted uses that enable time savings above $25\%$.}
\label{fig:bottlenecks-conditional}
\vspace{-3mm}
\end{figure}

\paragraph{Exposure analysis reveals bottlenecks.}
By directly accounting for how we predict LMs save worker's time, we can inspect the different factors that contribute to task exposure predictions. 
Our exposure estimates surfaces specific real-world \textit{bottlenecks} that limit AI-enabled time savings on a task-by-task basis. We summarize our findings by using an LM to categorize our exposure measure's produced justifications why an LM would/wouldn't save a worker time.
In \autoref{fig:bottlenecks-conditional}, we visualize this breakdown across different bottleneck categories.
As expected, physical interaction requirements are the most common bottleneck preventing LMs from saving worker time. Additionally, tasks requiring real-time monitoring, live interaction, privacy constraints, and proprietary systems account for a substantial share of tasks that are labeled as not exposed by our simulation-based exposure measure. Rubric-based measures, which assume a fixed list of capabilities, cannot separate tasks that are ``physically impossible'' from those that are ``feasible but bottlenecked on privacy or proprietary systems.''
On the other hand, the tasks with the greatest time savings often involve steps like drafting long-form text (\eg reports), drafting communication-related text (\eg emails), and generating code.

\paragraph{Comparing to rubric-based exposure}

To implement the rubric-based exposure measure we use the prompt and rubric from \citet{hosseini2026generative_entry_barriers} rather than from \citet{eloundou2024gpts}: the former updates the prompt used by the latter to reflect current LMs, and they are strongly correlated; \citet{hosseini2026generative_entry_barriers}  report an $R^2 = 0.76$ between the two measures.

\autoref{fig:exposure-intensity} compares exposure intensity across \exposureIntensityNumOcc occupations, which constitute $\exposureIntensityPropOcc$ of the occupations in O*NET. We exclude the remaining occupations due to data generation errors encountered when computing the simulation-based exposure. We compute exposure intensity using either our proposed simulation-based exposure or the reproduced rubric-based exposure. We call a task highly exposed if the exposure measure predicts an LM can save $\geq50\%$ of time, and moderately or highly exposed at the $\geq25\%$ threshold. At both exposure levels, simulation-based exposure marks a smaller fraction of each occupation's tasks as exposed to AI than rubric-based exposure does. The former grounds LM capabilities in real performance via simulated interactions with an LM, while the latter assumes that LMs attain a list of capabilities that, in practice, are likely not fully achieved (e.g., "Data analysis," "Business analytics," "Idea generation"). The simulation-based exposure measure suggests that $\percentOccupationsSaveHalfTasks$ of occupations have at least $50\%$ of their tasks at a medium or high AI exposure level, compared to $\percentOccupationsSaveHalfTasksHL$ under the rubric-based exposure measures. 
We posit that grounding in simulated interactions with an LM enables the simulation-based measure to better estimate exposure to \textit{current} LMs.
\section{Conclusion}
\label{sec:conclusion}
We introduce \projectname as an open-source evaluation suite for measuring language model capabilities on U.S. work categories.
By providing both benchmark-style and exposure-style measures, expanding coverage of U.S. work, and operating at the task, DWA, and occupational levels, we develop an array of tools for better measurement in this domain. 
We encourage future work to explore how these different measures predict downstream economic indicators like employment, wages, and productivity. 
In addition, future work can address the most fundamental limitations of our work, namely our technological scope being limited to current LM chatbots and our economic scope being limited to the current U.S. labor economy.
Overall, we contribute measurement infrastructure to support the broader research agenda and evidence base on the economics of frontier AI.
\section*{Acknowledgements}
We thank Abhishek Nagaraj, Amir Zeinali, Andy Haupt, Arjun Ramani, Arvind Narayanan, Avanika Narayan, Bharat Chandar, Dan Ho, Dawn Song, Divya Siddharth, Diyi Yang, Dylan Clement, Erik Brynjolfsson, Jacob Steinhardt, Jaime Sevilla, Joel Becker, Jon Saad-Falcon, Kris Gulati, Lukas Freund, Lukas Mann, Peter Cihon, Phil Trammell, Parker Whitfill, Rob Reich, Sam Manning, Sayash Kapoor, Tejal Patwardhan, Tom Cunningham, Yijia Shao, and Yifan Mai for helpful discussion.
We thank the Stanford Center for Research on Foundation Models (Stanford CRFM) and Stanford Institute for Human-Centered Artificial Intelligence (Stanford HAI) for funding.

\bibliographystyle{unsrtnat}
\bibliography{neurips_2026}

\clearpage
\appendix
\section{Real Data}
\label{app:real-data}

Given public user-chatbot conversations, we implement a pipeline that implements embedding-based retrieval followed by language model classification to map from conversations to detailed work activities (DWAs). 
Here we describe this pipeline in full detail.
Then we describe how we assess the pipeline and analyze alternatives.

\subsection{Pipeline implementation}
\label{app:real-pipeline-implementation}
\paragraph{Embedding.}
All of the embeddings used for the retrieved conversations are embedded as-is (i.e., concatenated turns, with each turn prefixed with ``Human:'' or ``Assistant''). For O*NET tasks, we ensemble two sets of embeddings: (1) one that just embeds the concatenated task description and occupation title; (2) averaged embeddings across synthetically generated conversations. We use OpenAI's \texttt{text-embedding-3-small} model.

\paragraph{Selecting promising DWAs.}
There are 2,087 total DWAs. As we expect the majority of DWAs to have few positive classifications, we first select the most promising DWAs. We do this by first performing a ``shallow'' retrieval with $k=10$ (looking at the top-10 most semantically similar conversations for each adjacent task). We then perform the rest of the classification pipeline and select the 200 DWAs with the most positive classifications.

\paragraph{Language model classification.}
Given the 200 DWAs we select, we retrieve the top 300 most similar conversations to perform pairwise classification for the resulting $200 \times 300$ pairs.

LMs are used to perform binary classification for a conversation-category pair, where the final output is whether that conversation belongs in that category. However, we may also preprocess the conversations or categories before passing them to the LM.

Although the final category that we care about are the DWAs, we can also classify against the lower-level task statement/occupations and also the higher-level IWAs. We classify against the lower-level task statements/occupations because the DWA titles by themselves are underspecified (e.g., DWAs alone don’t define the set of occupations to consider, but the adjacent task statements do) and we classify against the higher-level IWAs because it allows us to filter out conversation-category pairs with fewer inferences (since there are fewer IWAs than DWAs).

Finally, we also may preprocess the user requests or the category descriptions prior to add information that improves the specificity of the classification or remove information that may bias the language model classifier.

For the conversations, we perform up to three preprocessing steps:
\begin{enumerate}
    \item For all current configurations, only the user requests (and not the assistants’ responses) are given to the model. Assistant responses tend to be verbose, which uses unnecessary tokens. But also, LMs have trouble ignoring misleading context and we want LMs to focus on the user requests (since that’s what we’re evaluating models on), so including assistants’ responses tends to reduce accuracy as well.
    \item We may filter out turns irrelevant to the category. This is done for a similar reason to the above: LMs have trouble ignoring irrelevant context \& conversations often include turns like ``Human: can you speak Spanish?''
    \item In certain prompt configurations, we may summarize the user requests as well. This is also done for a similar reason as the above: user requests contain a lot of extraneous information \& summarization allows us to guide LMs to focus on the correct parts.
    
\end{enumerate}

We also preprocess the categories. DWA titles, IWA titles, and Task statements tend to be very brief (one sentence or less than one sentence), but the structure of the O*NET hierarchy includes information we can add: we use an LM to summarize e.g., adjacent DWAs (to add detail to the IWA titles), adjacent task statements (to add detail to the DWA titles) or contrast task statements for an occupation with other task statements that belong to that occupation (to add detail to the task statements).\\

\begin{table*}[t]
\centering
\small
\setlength{\tabcolsep}{4pt}
\renewcommand{\arraystretch}{1.2}
\begin{adjustbox}{max width=\textwidth}
\begin{tabularx}{\textwidth}{
    c c c
    >{\raggedright\arraybackslash}X
    >{\raggedright\arraybackslash}X
}
\toprule
\textbf{Step} & \textbf{Category} & \textbf{Model} & \textbf{Prompt preprocessing} & \textbf{Category preprocessing} \\
\midrule

(1) & IWA & \texttt{openai/gpt-4.1-nano}
& Remove assistant turns
& Add details to IWA titles
\\

(2) & DWA & \texttt{openai/gpt-4.1-nano}
& Remove assistant turns
& Add details to DWA titles
\\

(3) & Task & \texttt{openai/gpt-4.1-mini}
& Remove assistant turns
& Add details to task statements
\\

(4) & DWA & \texttt{openai/gpt-4.1-mini}
& Remove assistant turns, filter irrelevant turns and summarize user queries
& Add details to DWA titles
\\

(5) & Task & \texttt{openai/gpt-4.1-mini}
& Remove assistant turns, filter irrelevant turns and summarize user queries
& Add details to task statements
\\

(6) & Task & \texttt{openai/gpt-4.1-mini}
& Remove assistant turns, filter irrelevant turns and summarize user queries
& Add details to task statements
\\

\bottomrule
\end{tabularx}
\end{adjustbox}
\caption{Summary of language model classification steps used to categorize conversations into economic categories.}
\label{tab:classification-turns}
\end{table*}

In total, we ensemble across six turns of classification, which we summarize in \autoref{tab:classification-turns}.

Finally, we use an LM to filter out non-work and low-quality queries: for the former, we remove conversations that are obvious homework, coursework, or exam-style requests that the user wants completed; for the latter, we remove conversations that are unanswerable from the provided context, require capabilities a text-only LM does not have, or are otherwise illegitimate or joke-like rather than genuine work-related requests.

\subsection{Pipeline evaluation}
\label{app:real-pipeline-evaluation}

\begin{figure}[p]
    \centering
    \begin{subfigure}{0.49\linewidth}
        \centering
        \includegraphics[width=\linewidth]{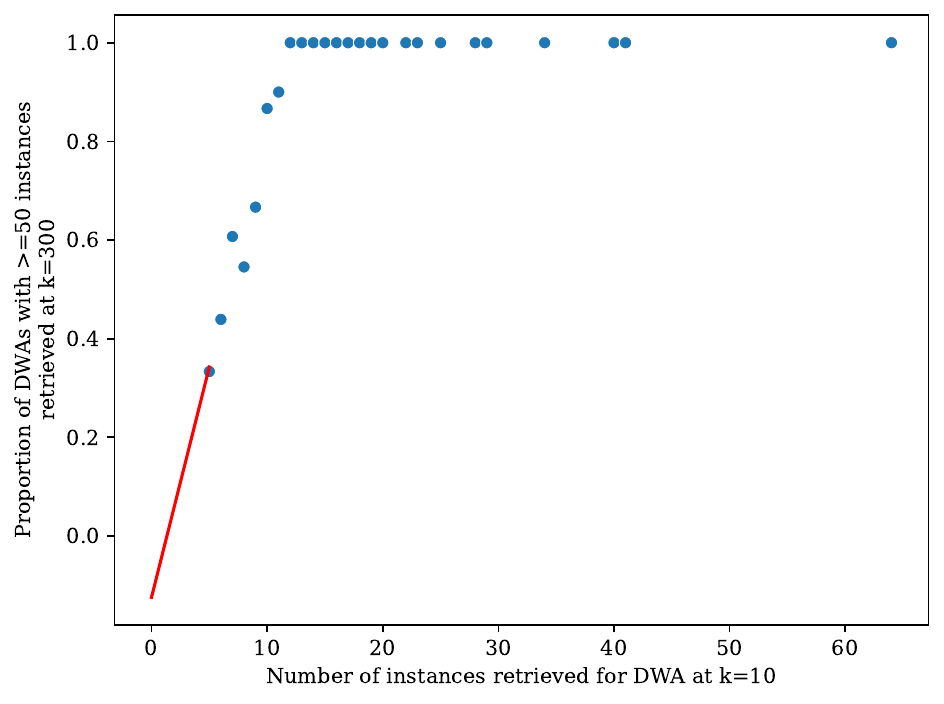}
        \caption{Number of relevant samples retrieved during the shallow-retrieval versus the full run.}
        \label{fig:shallow-trial}
    \end{subfigure}
    \hfill
    \begin{subfigure}{0.49\linewidth}
        \centering
        \includegraphics[width=\linewidth]{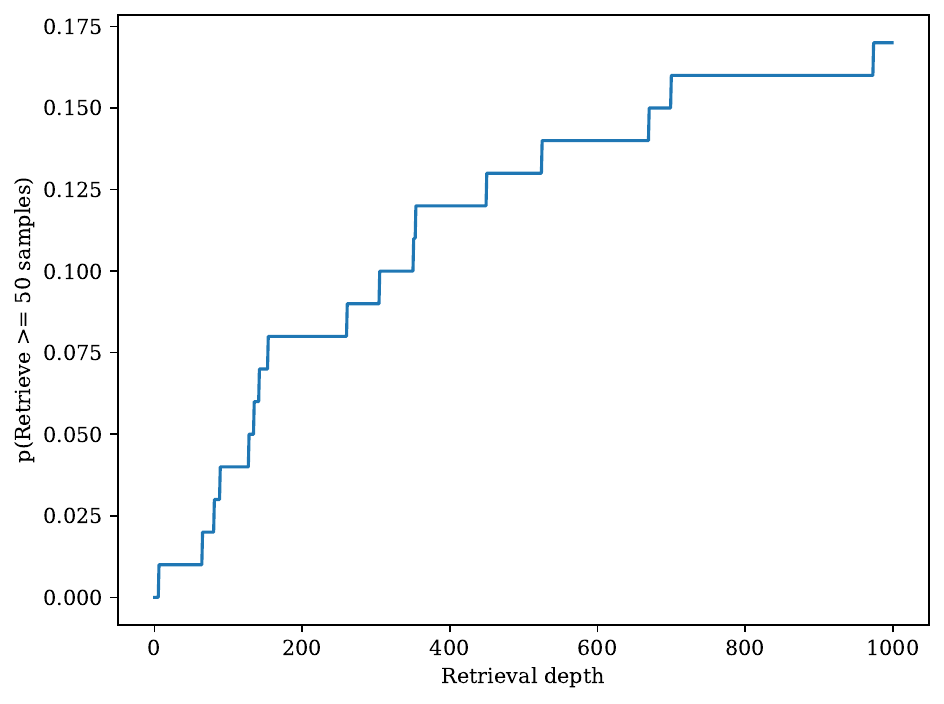}
        \caption{Number of relevant samples retrieved versus retrieval depth.}
        \label{fig:depth-trials}
    \end{subfigure}
    \caption{\textbf{Results from analyzing the coverage of the retrieval pipeline.}}
    \label{fig:retrieval-trials}
\end{figure}

There are three sources of error where we may fail to retrieve at least 50 samples for a DWA despite at least 50 samples existing in real data.

First, we only perform retrieval on the 200 most promising DWAs and may miss samples from the remaining 1,887 DWAs. We select the promising DWAs based on the number of relevant samples when looking at only the top-10 most semantically similar samples. As such, by looking at the relationship between the number of relevant samples retrieved during this ``shallow'' trial to the rate at which we successfully find fifty relevant samples in the final run (looking at the top-300 most semantically similar samples), we can estimate the number of DWAs we missed by only looking at the 200 most promising ones rather than the full set. Specifically, we observe a linear relationship between the number relevant samples retrieved in the shallow trial and the probability we retrieve at least fifty samples in the full run (\autoref{fig:shallow-trial}) and estimate that there are \dwasOutsideOfSelected DWAs that we missed by only looking at the 200 most promising DWAs rather than the full set of 2,087.

Second, we only perform classification looking at only top-300 most semantically similar conversations and may miss non-semantically similar conversations. We estimate the error from embedding-retrieval by running the pipeline on subsets of DWAs and comparing the number of relevant samples retrieved at different depths: 300, 1,000, and 10,000 (\autoref{fig:depth-trials}). We estimate that the probability we would have retrieved at least fifty relevant samples at depth 1,000 conditional on \textit{not} retrieving at least fifty samples at depth 300 is \pRetrieveAtThousandButNotThreeHundred and the probability we would have retrieved at least fifty relevant samples at depth 10,000 conditional on \textit{not} retrieving at least fifty samples at depth 300 is \pRetrieveAtTenThousandButNotThreeHundred. This means that, for the 200 most promising DWAs, there are an additional \exprAdditionalDWAsFromGreaterDepth we could have gotten by retrieving at a 33.3x greater depth and \exprAdditionalDWAsFromGreaterBreadth for the DWAs that we did not try performing retrieval on, resulting in a total of \exprAllMissedDWAs additional DWAs.

Finally, the classification pipeline prioritizes high precision and may have false negatives. We sample 50 conversation-category pairs for which the pipeline classified as not relevant, and find that we agree with the prediction 73.4\% of the time.    

\section{Synthetic Data}
\label{app:synthetic-data}

We find that simple prompts tend to result in simplistic generations (e.g., just rephrasings of the task description). Instead, we use a prompt where the LM roleplays as a worker sharing how they use AI chatbots to assist with a given task.

Specifically, for each O*NET task and occupation, we first generate a worker ``persona'' consisting of the company/organization, the years of experience, specific job title, etc. We then use \texttt{GPT-5-mini} to roleplay as this worker persona. The user roleplays as an ``interviewer'' asking the ``worker'' to send them a prompt that helped save them time on the task. Crucially, we condition in this prompt the amount of time-savings that prompt provides for this task. This lets us generate prompts with varying degrees of ``comprehensiveness'': we can generate a complex prompt that performs the entire task if we ask the ``worker'' to send a prompt with 90-100\% time-savings; we can also generate a simpler prompt with 1\% time-savings.

Ideally, we'd like to generate as comprehensive of a prompt as possible, but an LM may be willing to hallucinate (or ignore certain constraints) to generate a prompt with high time-savings. So, we wrap this in a loop with additional prompts (using a more capable \texttt{GPT-5.2}) to verify that (1) the prompt reflects things that someone of this occupation would perform and (2) that the prompt is actually answerable. The full pipeline for synthetic generation then consists of trying to generate, prompting with the highest time-savings (90-100\%) then lowering this time-savings amount until we get a synthetic prompt that passes the verifiers. Finally, for tasks that do not have passing synthetic prompts across all time-savings values, we perform additional attempts by restarting the pipeline from the beginning.

\section{Benchmarks}
\label{app:benchmarks}

We report results on our benchmarks along with comparing the costs of producing our benchmarks with the costs of producing OpenAI's GPDval \citep{patwardhan2025gdpval}.

\subsection{Benchmark generation costs}
\label{app:gdpval_cost_calc}
Since GDPval is the state-of-the-art economic evaluation prior to our work, and perhaps even after our work given the data quality, we compare our benchmark generation procedure to their dataset.
We estimate that generating synthetic data such that we have at least 50 instances for each of the 43 occupations costs less than \$1,000.
In contrast, generating (and validating) worker queries for OpenAI's GDPval to have at least 30 instances costs at least \$500,000 according to our estimate.

To estimate the overall query generation cost, we study the exact GDPval benchmark design process, the selected occupations, and the associated wages.
Assuming that workers involved in GDPval were paid their standard wage per unit time, we get an estimate of $\$566{,}296$ via a back-of-the-envelope calculation.
GDPVal reports that each task was reviewed by an average of 5 people during creation, and that the mean review time for a model's output on a task was 109 minutes. We use the latter as a proxy for per-reviewer task-creation review time, noting that these are conceptually distinct activities. We exclude the time spent initially drafting or editing tasks, as GDPVal does not report values for these activities; including them would scale our estimate upward. We then compute per-occupation cost as $30 \text{ tasks} \times 5 \text{ reviews} \times (109/60) \text{ hours} \times \text{mean hourly wage for occupation}$, and sum across the 44 occupations covered by GDPVal. We rate our confidence in this estimate as low-to-medium, and emphasize that it likely represents a lower bound on the true cost.

\begin{figure}[p]
    \centering
    \includegraphics[width=\linewidth]{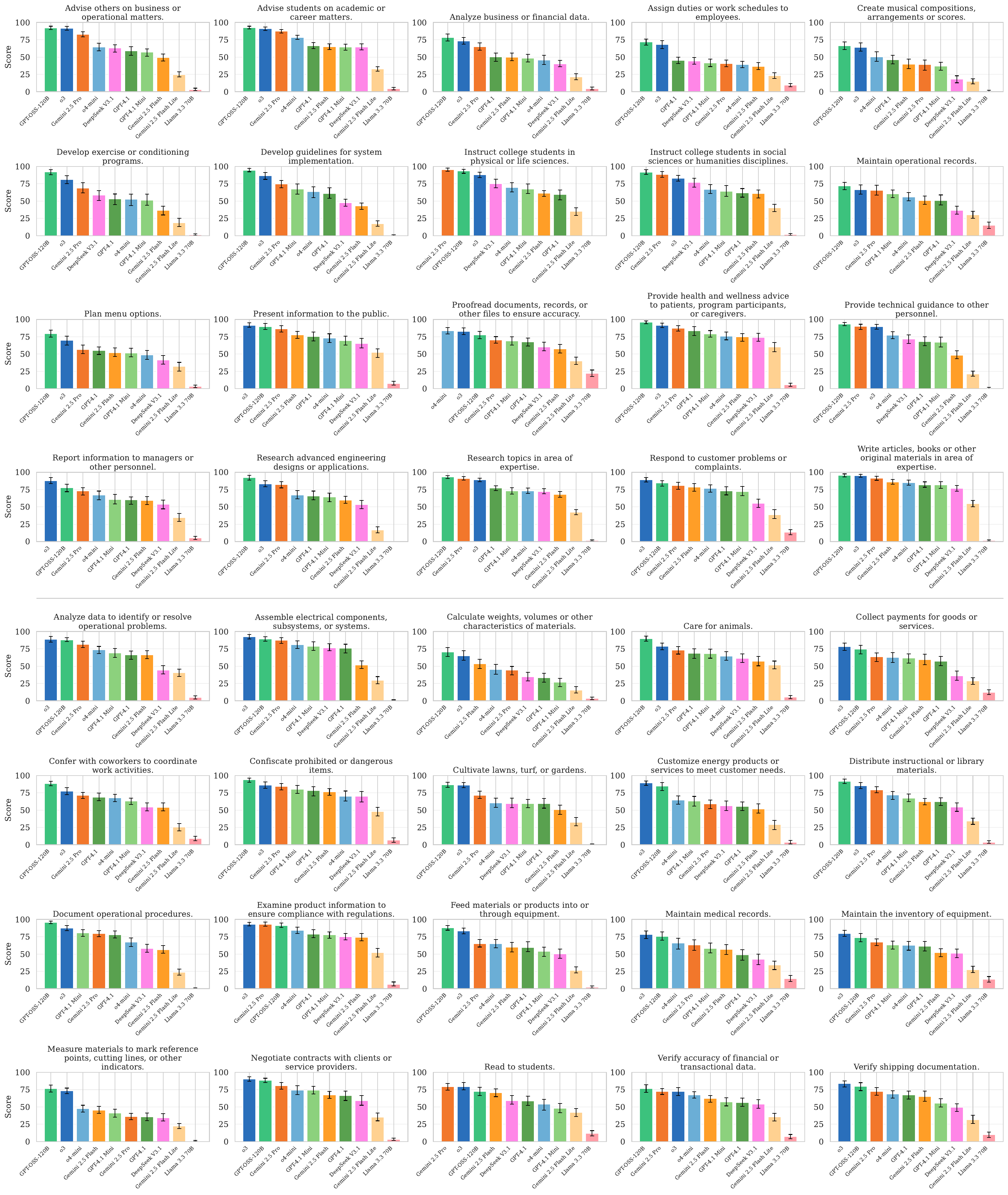}
    \caption{\textbf{DWA-level synthetic benchmark results.}
    The top-half correspond to DWAs also found in retrieved data. The bottom-half correspond to DWAs not found in retrieved data.
    }
    \label{fig:synthetic-DWA}
\end{figure}

\begin{figure}[p]
    \centering
    \includegraphics[width=\linewidth]{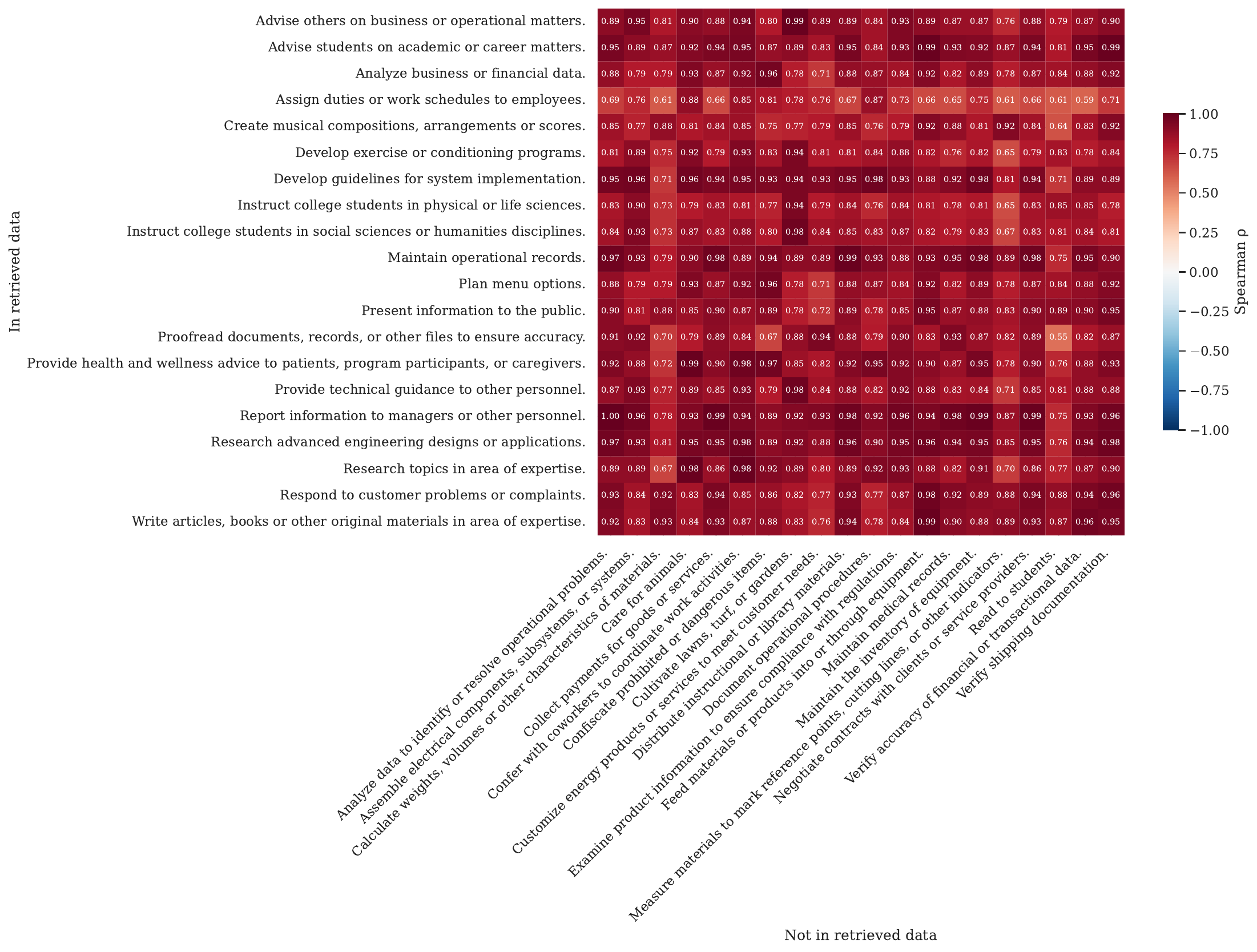}
    \caption{\textbf{Spearman correlation between leaderboards where the DWA is found in real usage versus not found in real usage.}
    }
    \label{fig:use-vs-nonuse-benchmarks}
\end{figure}

\begin{figure}[p]
    \centering
    \includegraphics[width=0.9\linewidth]{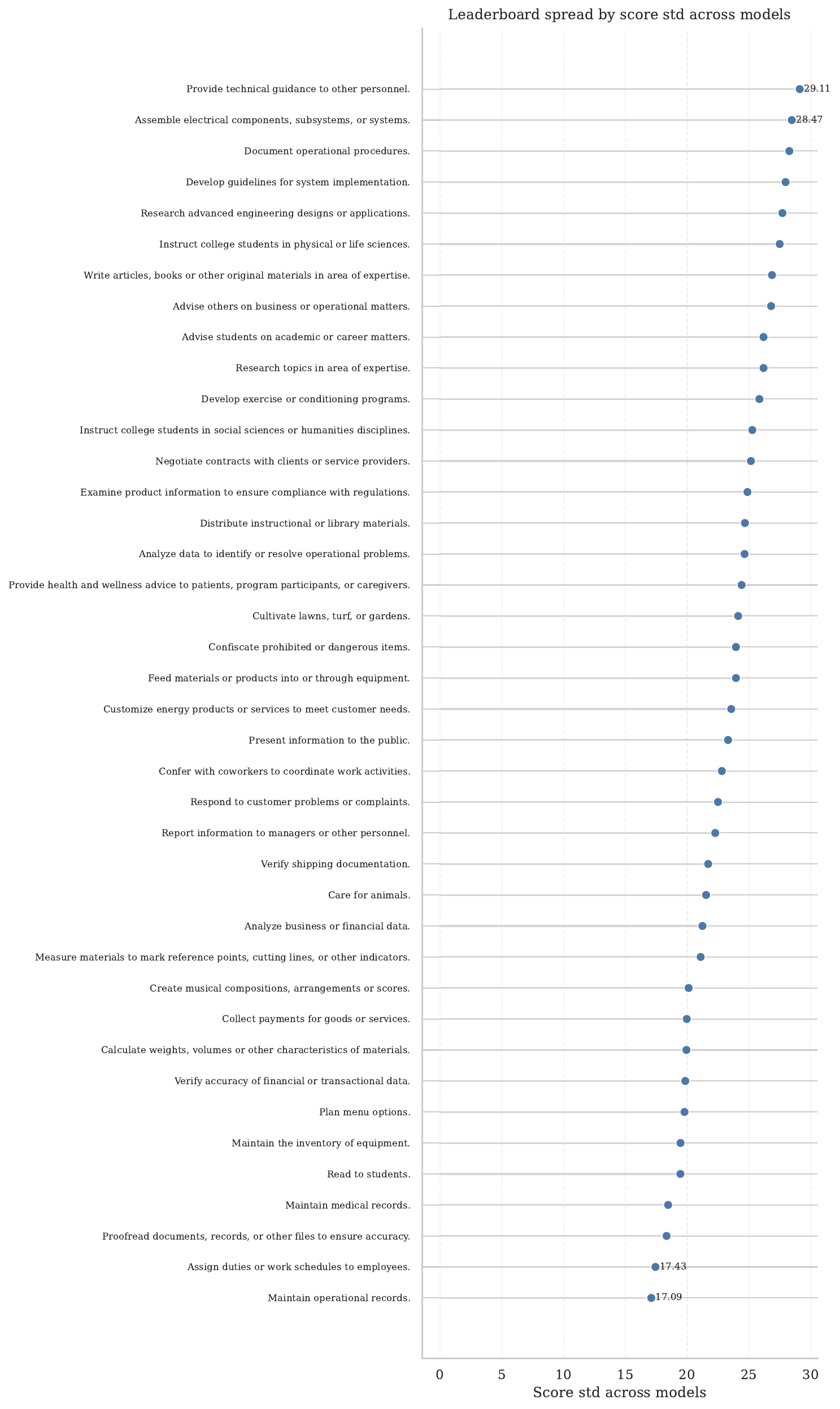}
    \caption{\textbf{Variance in DWA-level synthetic benchmark results.}
    We rank DWAs by the variance they induce in model performance for the 40 DWA-level synthetic benchmarks.
    }
    \label{fig:synthetic-DWA-variance}
\end{figure}

\begin{figure}[p]
    \centering
    \includegraphics[width=\linewidth]{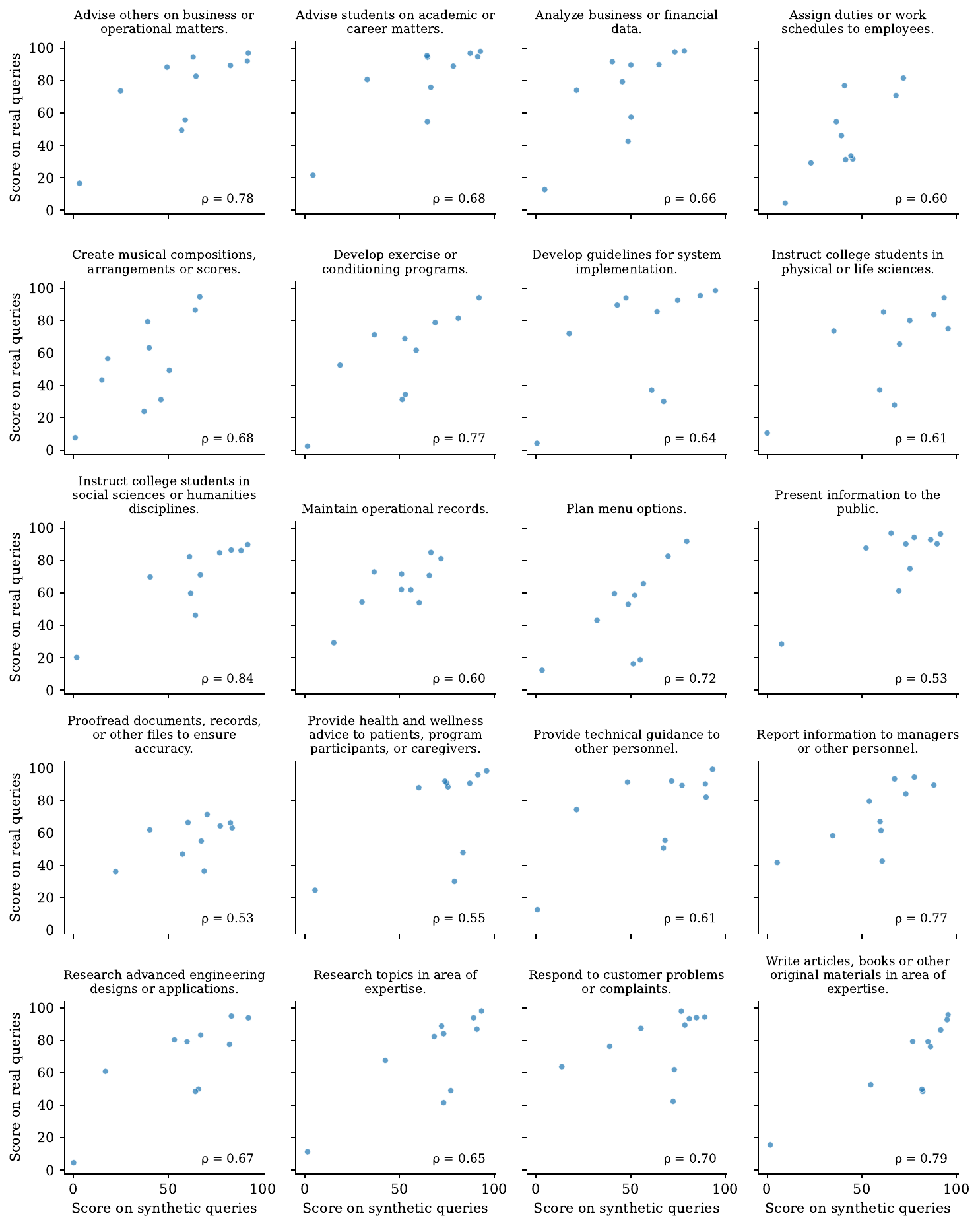}
    \caption{\textbf{DWA-level synthetic benchmark results compared to real benchmark results.}
    }
    \label{fig:real-vs-synthetic-benchmarks}
\end{figure}

\begin{figure}[p]
    \centering
    \includegraphics[width=0.8\linewidth]{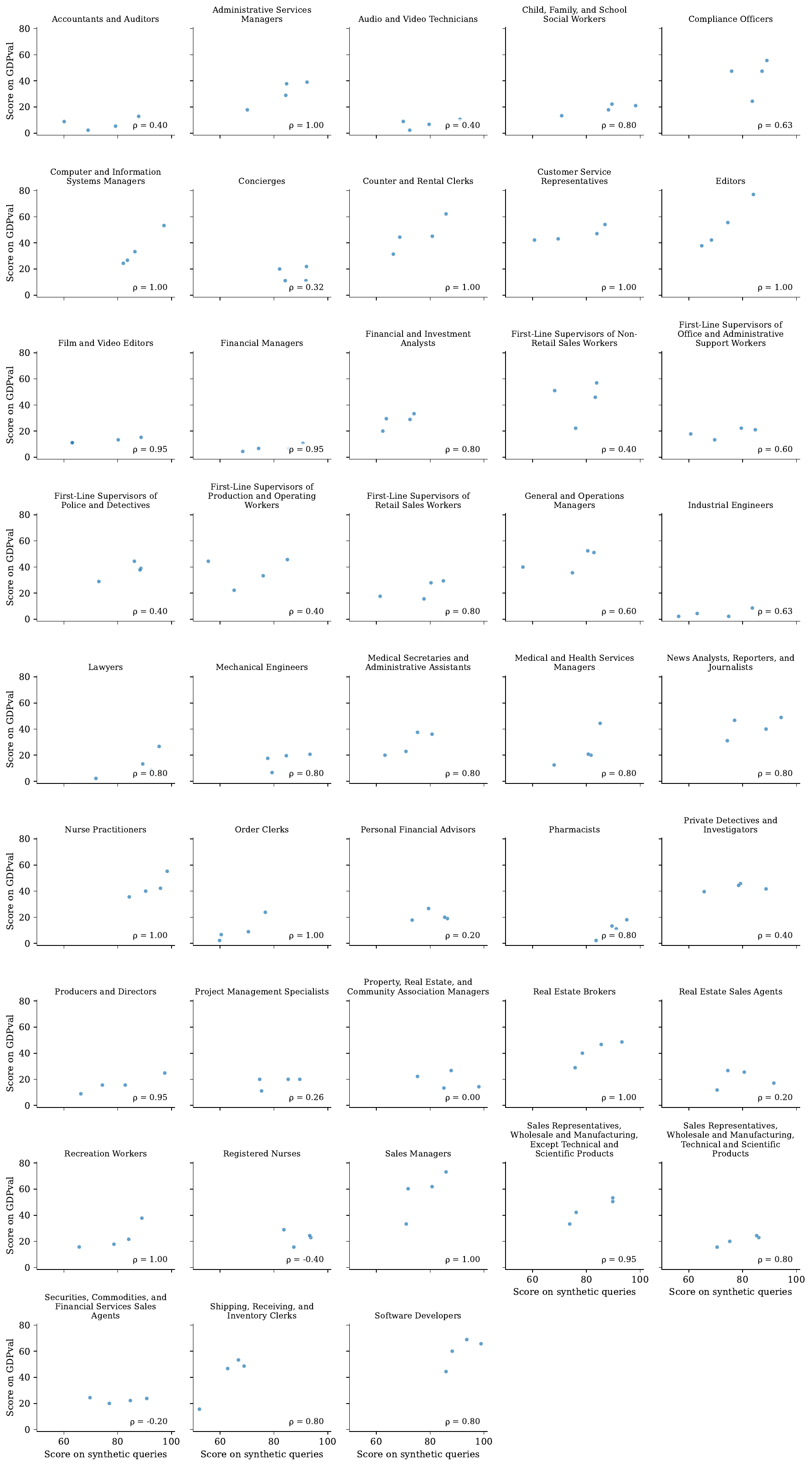}
    \caption{\textbf{Occupation-level synthetic benchmark results compared to GDPval benchmark results.}
    }
    \label{fig:gdpval-vs-synthetic-benchmarks}
\end{figure}

\begin{figure}[p]
    \centering
    \includegraphics[width=\linewidth]{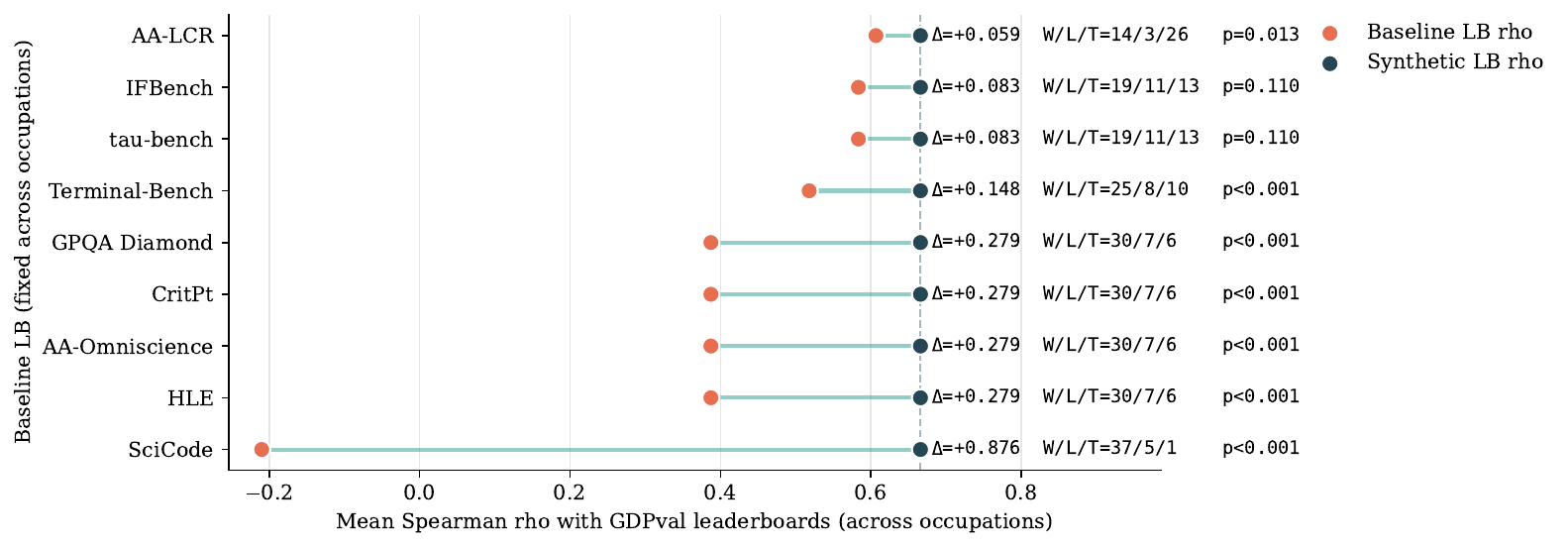}
    \caption{\textbf{Correlation between synthetic benchmark results and GDPval benchmark results versus generic capabilities benchmarks.}
    }
    \label{fig:predicting-GDPval}
\end{figure}

\section{Exposure}
\label{app:exposure}

\subsection{Predicting exposure through simulation}
\label{app:detailed_exposure_description}

\begin{table}[h]
\centering
\begin{tabular}{>{\raggedright\arraybackslash}m{11cm} c c}
\toprule
\textbf{Step} & \textbf{Orig.} & \textbf{Saved} \\
\midrule
\color{gray} The respondent needs to run a short intake conversation with the product manager to clarify acceptance criteria and update the ticket with unambiguous inputs. & \color{gray} 30 & \color{gray} 10 \\ \addlinespace[3mm]
The respondent needs to review the existing repository and system documentation to locate the reconciliation module, inspect schemas, and note compatibility and performance constraints. & 45 & 10 \\ \addlinespace[3mm]
The respondent needs to design a high-level workflow diagram showing ingest, normalize, match, apply business rules, update the ledger, and escalate, including operational components like idempotency and dead-letter queues. & 60 & 25 \\ \addlinespace[3mm]
The respondent needs to define data models, API/interface definitions, and example JSON payloads and draft the \texttt{reconciliation\_history} database table schema for downstream implementers. & 40 & 17 \\ \addlinespace[3mm]
The respondent needs to write the core algorithm pseudocode for the matcher, enumerate edge cases, and design concurrency control. & 45 & 25 \\ \addlinespace[3mm]
\color{gray} The respondent needs to break the work into implementation subtasks, estimate effort for a two-engineer sprint plan, and list dependencies and acceptance criteria. & \color{gray} 20 & \color{gray} 10 \\ \addlinespace[3mm]
\color{gray} The respondent needs to run a peer review walkthrough of the design with engineers and product, collect feedback, and record action items. & \color{gray} 25 & \color{gray} 8 \\ \addlinespace[3mm]
\color{gray} The respondent needs to incorporate review feedback, finalize the design document with acceptance tests and diagrams, and publish the doc linked to the implementation tickets. & \color{gray} 25 & \color{gray} 10 \\
\bottomrule
\end{tabular}
\vspace{2mm}
\caption{An example step-by-step breakdown produced through simulation for the task ``Write, analyze, review, and rewrite programs, using workflow chart and diagram, and applying knowledge of computer capabilities, subject matter, and symbolic logic.'' corresponding to the occupation of ``Computer Programmers''. \textit{Orig.} corresponds to the original time (in minutes) that the step is predicted to take and \textit{Saved} corresponds to the amount of time saved. The lines in grey are steps that we filter out for being irrelevant to the target task.}
\label{tab:example-steps-breakdown}
\end{table}

For each task, we use an LM to simulate an interview between a labor economist and a worker, where the LM plays the role of the worker. For each occupation and task the LM responds in several turns to produce a step-by-step breakdown of the chatbot-enabled time-savings. Specifically, it:
\begin{enumerate}
    \item Generates a more detailed background to ground subsequent responses, including job title, employer, and years of experience.
    \item Generates a more detailed description of the work task.
    \item Generates a step-by-step breakdown of how the work task is performed (including the amount of time each step takes).
    \item Given a synthetically generated prompt for the work task (see \autoref{app:synthetic-data}) and an LM response to the prompt, walks through the previously generated step-by-step breakdown and notes instances when a chatbot with the demonstrated capabilities would or would not be useful.
    \item Using these responses, provide an estimate of the amount of time-savings a chatbot would enable across the steps.
    \item Output the previous step-by-step time-savings breakdown in a parseable format.
\end{enumerate}

Finally, as the LM has a tendency to include steps irrelevant to the provided task (or steps that fit better with other tasks), we follow with a post-processing step where we use an LM to filter out such steps given the original model responses and the full list of tasks for the occupation. An example of the final output, containing a step-by-step breakdown of relevant time-savings, can be found in \autoref{tab:example-steps-breakdown}.

\subsection{Understanding low-exposure tasks with at least some Claude usage and vice versa}
\label{app:low_exposure_nonzero_usage}

\begin{figure}[h]
    \centering
    \begin{subfigure}[b]{0.49\textwidth}
        \centering
        \includegraphics[width=\textwidth]{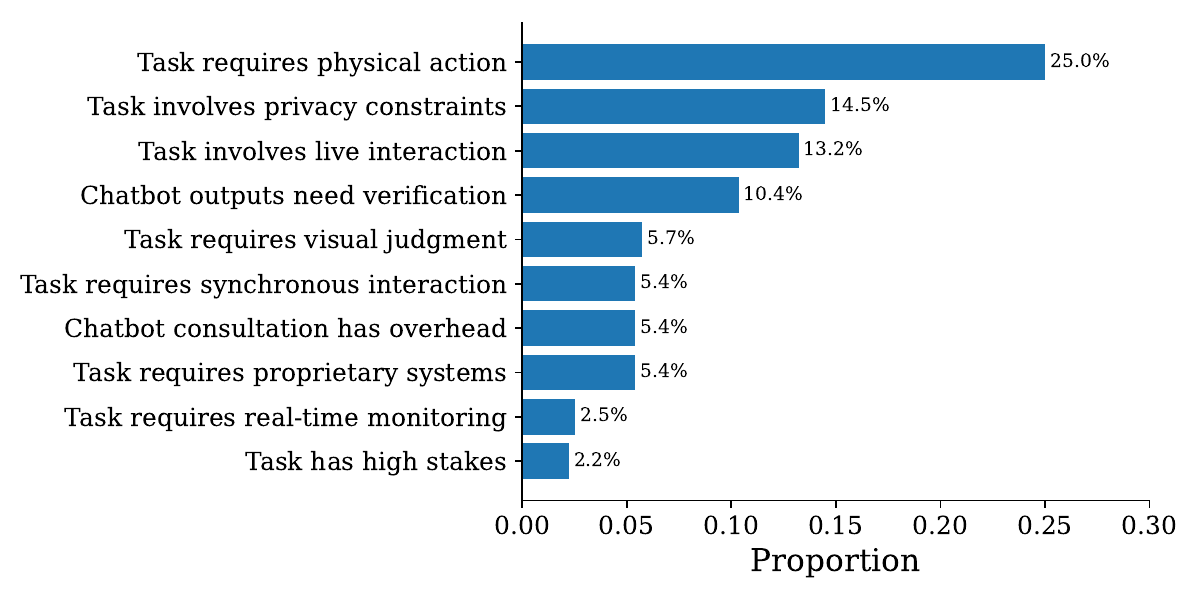}
        \caption{Bottlenecks driving predicted time savings below $25\%$ for tasks that have substantial Claude usage.}
        \label{fig:subfig-a}
    \end{subfigure}
    \hfill
    \begin{subfigure}[b]{0.49\textwidth}
        \centering
        \includegraphics[width=\textwidth]{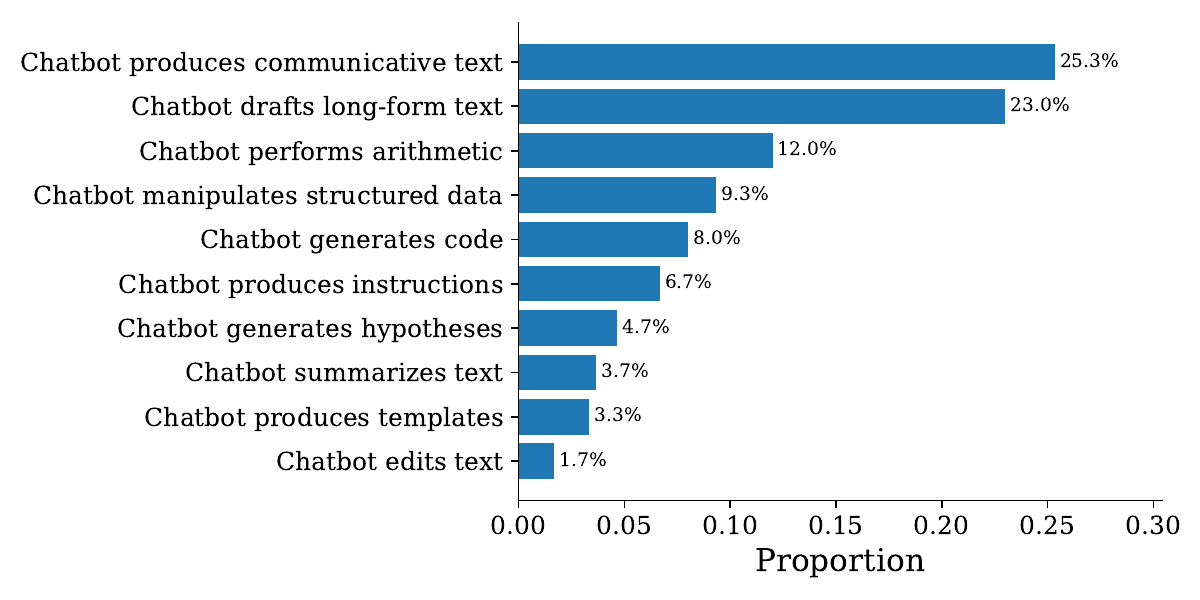}
        \caption{Factors driving predicted time savings above $25\%$ for tasks that have minimal Claude usage.}
        \label{fig:subfig-b}
    \end{subfigure}
    \caption{\textbf{The factors that result in the simulation-based exposure measure to predict substantial/non-substantial time-savings when Claude usage data suggests the opposite.}}
    \label{fig:bottlenecks-conditional-usage}
\end{figure}

Figure \ref{fig:bottlenecks-conditional-usage} complements Figure \ref{fig:exposure-usage}, showing the reasons why some tasks that see Claude usage are predicted to have low exposure by our simulation-based exposure measure, and why some tasks that see minimal Claude usage are predicted to have high exposure.  

\subsection{Skill Importance}

Following \cite{eloundou2024gpts}, we next examine how our exposure estimates relate to the skills that occupations rely on most heavily. The O*NET taxonomy defines ten Basic Skills across two categories:
\begin{itemize}
    \setlength{\itemsep}{0pt}
    \item \textbf{Content skills:} \textit{Reading Comprehension} (understanding written work documents), \textit{Active Listening} (attentive listening without interruption), \textit{Writing} (effective written communication), \textit{Speaking} (effective oral communication), \textit{Mathematics} (using math to solve problems), \textit{Science} (using scientific methods to solve problems).
    \item \textbf{Process skills:} \textit{Critical Thinking} (logic and reasoning to evaluate alternatives), \textit{Active Learning} (applying new information to problem-solving), \textit{Learning Strategies} (selecting appropriate methods for learning or teaching), \textit{Monitoring} (assessing performance to drive improvement).
\end{itemize}

O*NET provides labels for the importance of each skill for each occupation. We rescale each skill's per-occupation importance score, and then regress our exposure measures on these rescaled importance scores to characterize which skills are most strongly associated with AI exposure. 

We run this regression analysis to predict the simulation-based exposure measures and, separately, the rubric-based exposure measures. At the moderate-or-higher exposure threshold, both methods agree closely: language-intensive skills (Writing, Reading Comprehension, Critical Thinking, Speaking) are the strongest correlates of exposure, and the regressions explain a similar share of variance ($R^2 = 0.718$ for simulation-based vs.\ $0.738$ for rubric-based). At the high-exposure threshold, however, the two methods diverge sharply: rubric-based exposure remains well-explained by skill importance ($R^2 = 0.528$), with Programming emerging as the dominant correlate, whereas simulation-based exposure is only weakly predicted by O*NET skills ($R^2 = 0.085$) and exhibits no single dominant skill. This divergence suggests that high exposure under our method is not concentrated in occupations defined by any one skill, but is instead distributed more diffusely across occupational profiles.

\begin{table}[h]
\centering
\begin{tabular}{lcccc}
\toprule
 & \multicolumn{2}{c}{\textbf{Moderate + High Exposure}} & \multicolumn{2}{c}{\textbf{High Exposure}} \\
\cmidrule(lr){2-3} \cmidrule(lr){4-5}
 & \textit{Simulation Based} & \textit{Rubric Based} & \textit{Simulation Based} & \textit{Rubric Based} \\
\midrule
$R^2$        & 0.718 & 0.738 & 0.085 & 0.528 \\
\midrule
\multicolumn{5}{l}{\textit{Top correlates}} \\
Writing                & $+0.75$ & $+0.76$ & $+0.21$ & $+0.32$ \\
Reading Comp.  & $+0.75$ & $+0.78$ & $+0.21$ & $+0.37$ \\
Critical Thinking      & $+0.67$ & $+0.64$ & $+0.17$ & $+0.20$ \\
Speaking               & $+0.66$ & $+0.68$ & $+0.20$ & $+0.24$ \\
Active Learning        & $+0.66$ & $+0.63$ & $+0.20$ & $+0.19$ \\
Active Listening       & $+0.64$ & $+0.69$ & $+0.16$ & $+0.28$ \\
Programming            & $+0.56$ & $+0.58$ & $+0.09$ & $+0.52$ \\
Learning Strategies    & $+0.56$ & $+0.54$ & $+0.21$ & $+0.08$ \\
Mathematics            & $+0.53$ & $+0.53$ & $+0.10$ & $+0.32$ \\
Monitoring             & $+0.37$ & $+0.39$ & $+0.14$ & $-0.07$ \\
Science                & $+0.29$ & $+0.34$ & $+0.01$ & $-0.04$ \\
\bottomrule
\end{tabular}
\vspace{2mm}
\caption{Correlations between O*NET Basic Skill importance and predicted AI exposure, under simulation-based and rubric-based measures, at moderate-or-higher and high exposure thresholds. $r$ is the Pearson correlation between skill importance and exposure.}
\label{tab:skill-exposure}
\end{table}

\subsection{Barriers to Entry}
\begin{figure}[ht]
    \centering
    \includegraphics[width=\textwidth]{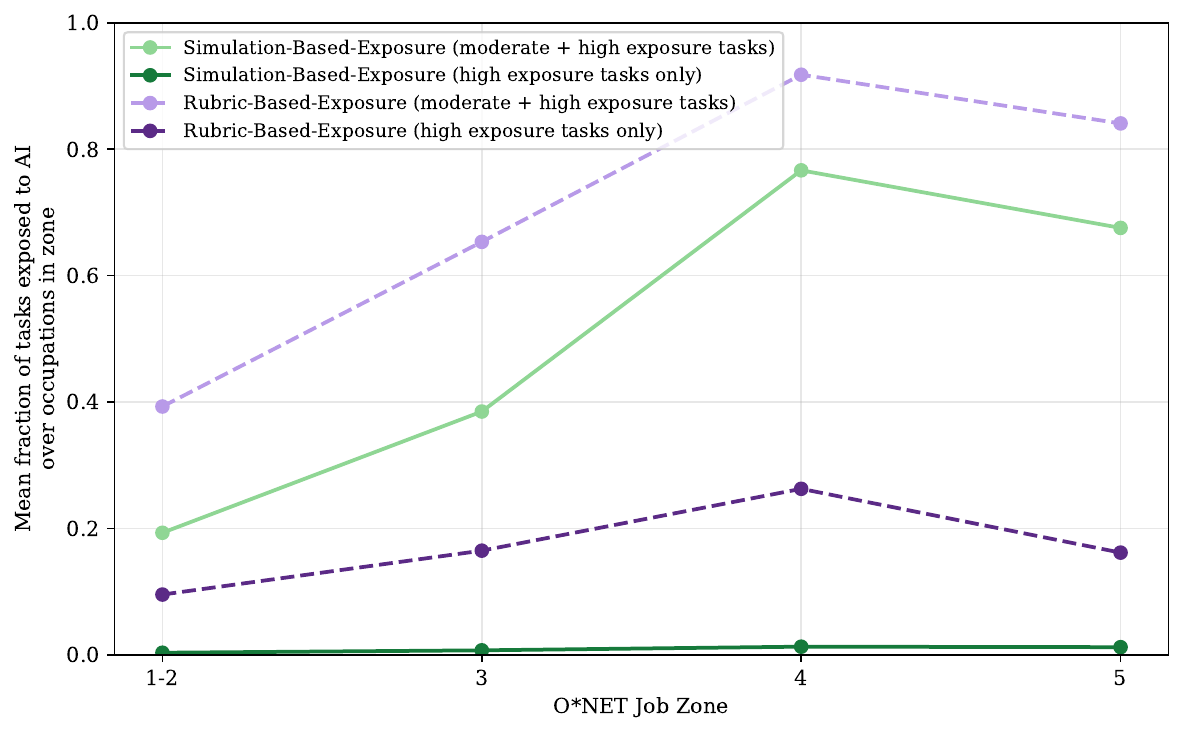}
    \caption{\textbf{Predicted exposure by O*NET Job Zone, under simulation-based and rubric-based exposure measures.} Job Zones group occupations by the education, experience, and training required for entry, ranging from Zone 1--2 (under one year of preparation) to Zone 5 (4+ years). Both measures exhibit the same pattern: exposure rises from Job Zone 1 through Job Zone 4 and then plateaus or declines at Job Zone 5.}
    \label{fig:exposure-by-jobzone}
\end{figure}

Following \cite{eloundou2024gpts}, we ask whether predicted exposure varies systematically with the preparation an occupation requires. To operationalize preparation, we use O*NET's Job Zone classification, which assigns each occupation to one of five tiers based on the education, prior experience, and on-the-job training typically needed to enter the role. Job Zone 1--2 occupations require minimal preparation (under one year), while Job Zone 5 occupations require four or more years. Plotting simulation-based and rubric-based exposure measures against Job Zone reveals the same pattern for both exposure measures: exposure rises from Job Zone 1-2 through Job Zone 4 and then plateaus or declines at Job Zone 5. Figure \ref{fig:exposure-by-jobzone} illustrates this result.

\end{document}